\documentclass[a4paper,10pt]{article}
    \usepackage{amsmath,amsfonts,amsthm,dsfont,amssymb,lipsum}
    \usepackage[blocks]{authblk}
    \newenvironment{keywords}{\noindent\textbf{Keywords:}}{}

    \newcommand{\email}[1]{\texttt{\small #1}}
	\newcommand{\orcid}[3]{\href{https://orcid.org/#1}{\includegraphics[width=7pt]{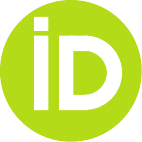}}}

    \usepackage{fancyhdr}

\usepackage{marvosym}
\usepackage{amsfonts}
\usepackage{amsmath}

\usepackage{hyperref}
\usepackage{color}
\usepackage{graphicx}
\usepackage{epsfig}
\usepackage[utf8]{inputenc}

\usepackage{sidecap}

\usepackage{wrapfig,lipsum,booktabs}

\hypersetup{colorlinks, citecolor=blue, linkcolor=red, urlcolor=blue}

\usepackage[title]{appendix}

\begin{document}
            \title{Relaxation dynamics of an unlike spin pair system
            \thanks{Corresponding author A. Consuelo-Leal }
            }

    \author[1]{A. Consuelo-Leal \footnote{E-mail: \email{adrianeleal25@gmail.com}}}  
    \author[1]{Hugo D. Fern\'{a}ndez Sare} 
    \author[2] {R. Auccaise} 
    \affil[1]{Departamento de Matem\'{a}tica, Universidade Federal de Juiz de Fora, Bairro S\~{a}o Pedro, 36036-900 Juiz de Fora, Minas Gerais, Brasil. }
    \affil[2]{Departamento de F\'{i}sica, Universidade Estadual de Ponta Grossa, Av. General Carlos Cavalcanti 4748, 84030-900 Ponta Grossa, Paran\'{a}, Brasil.}

    \date{(Received on 26 June 2023, Accepted on 16 October 2023)} 

    \newcommand\shorttitle{Relaxation dynamics of an unlike spin pair system}
    \newcommand\authors{A. Consuelo-Leal, Hugo D. Fern\'{a}ndez Sare and R. Auccaise.}

\maketitle

    \begin{abstract}
    Redfield master equation was applied to study the dynamics of an ensemble of interacting pairs of unlike spins at room temperature. This spin quantum system is a workbench quantum model to analyze the relaxation dynamics of a heteronuclear two-level spin system interacting by a pure dipole-dipole coupling. Expressions for the density matrix elements and their relaxation rate constants of each coherence order were computed. In addition, the solutions were evaluated considering three initial quantum states, and the theoretical predictions, such as multi-exponential evolutions and enhancement, are behaviors that the solutions preserve and agree with previous studies performed for magnetization time evolutions. Moreover, the solutions computed to predict the dynamics of the longitudinal magnetization avoid the disagreement reported by I. Solomon.
    \end{abstract}



\section{Introduction}
\label{sec:Introduction}

Dipole-dipole interaction is an essential interaction between two or more nuclei at inter-nuclear and intramolecular distances. Its applicability is checked in proteins and biology \cite{cavanagh2007Book,Fa2021}, animal navigation \cite{gauger2011}, organic compounds (chemistry) \cite{hausser1968,muller-warmuth1983}, polar molecules on atom physics \cite{yan2013}, the oil industry, and many others \cite{noggle1971Book,cui2005Book}. One of the main approaches to exploring the dipole-dipole interaction is the establishment of the relaxation dynamics of the particles. Based on time evolutions, if any spin system is driven into any excited state using whatever method, it can be monitored, measuring any observable physical quantity until it reaches its stationary stage. 

The pioneering work on relaxation was introduced by Bloembergen, Purcell, and Pound \cite{bloembergen1948}. Some years later, Solomon \cite{solomon1955} studied a like and an unlike spin pair system’s relaxation coupled by dipole-dipole interaction. The dynamics of both systems were used as a starting point to introduce the master equation approach reported by Redfield \cite{redfield1957} and Bloch \cite{bloch1957} independently. Indeed, there are some cases in which it is not necessary to develop a deep mathematical effort to solve the linear system of differential equations to know the characteristic relaxation rate constants of the nuclei. The procedure is well-detailed in many textbooks  \cite{abragam1994Book,levitt2001Book,kruk2016Book}. It can be verified for a nuclear homogeneous spin system or a like spin pair system (see Sec. $4.2$ of Ref. \cite{mcconnell1987Book})). If Redfield theory is applied to study the dynamics of the like spin system using the density matrix elements notation, then Solomons and Redfield approaches independently generate the same relaxation rate constants expressions for the transverse and longitudinal magnetization. However, the same analytical mathematical procedure is not an easy task, and in the following sections, we explain and detail the procedures and discuss some applications.

In the present letter, we introduce some mathematical methods and discuss physical arguments that allow us to suggest an analytical solution for an unlike spin pair system coupled weakly to an environment. Also, we highlight the main differences between Solomons and Redfields approaches. The paper is structured as follows: In Sect. \ref{sec:DescriptionSpinSystem}, we begin with a theoretical introduction of the spin system, the Redfield approach, and a detailed description of the solutions. Subsequently, in Sect. \ref{sec:Applications}, some applications are detailed, considering the time evolution of longitudinal magnetization at three initial quantum states. Then, Sect. \ref{sec:Discussions} opened a discussion to contextualize the solutions introduced in this paper and compare them with those previously computed by I. Solomon. In Sect. \ref{sec:Conclusions} has resumed our conclusions.

\section{Description of the nuclear spin system}
\label{sec:DescriptionSpinSystem} 

Let us  briefly introduce an unlike nuclear spin pair system. Consider two-level spins interacting with each other, denoted by $I$ and $S$ species, respectively. The total energy of the spin pair is the contribution of the Zeeman energy and the dipole-dipole energy interaction. The Zeeman energy represents the magnetic moment of each species interacting with a strong static magnetic field $B_{0}$ along the $z$-axis of a reference frame $XYZ$, $\mathbf{B}_{0}=\left(0,0, B_{0}\right)$. This energy contribution is expressed by $-\hbar \gamma ^{I} {\mathbf{I}} \cdot \mathbf{B}_{0} =-\hbar\omega ^{I}{\mathbf{I}} _{z} $, and similarly for the specie $S$, where ${\mathbf{I}}=\left( {\mathbf{I}}_{x},{\mathbf{I}}_{y}, {\mathbf{I}}_{z}\right) $ and ${\mathbf{S}}=\left( {\mathbf{S}} _{x},{\mathbf{S}}_{y},{\mathbf{S}}_{z}\right) $ mean the vectors of spin angular momentum operators,  $\omega ^{I,S}=\gamma^{I,S} B_{0}$ are the Larmor frequencies of both species and $\hbar$ represents the reduced Planck's constant.

Therefore, the Zeeman energy is mathematically quantified by the Hamiltonian operator denoted by ${\mathcal{H}}_{0}=-\hbar \omega ^{I}{\mathbf{I}}_{z}-\hbar \omega ^{S}{\mathbf{S}}_{z}$. This energy contribution establishes the characteristic energies such that from the fundamentals of Quantum Mechanics, a basis of quantum states that establishes the Hilbert space $\mathbb{H}$ for this spin system defines the set $\left\{ \left\vert \uparrow \uparrow \right\rangle ,\left\vert \uparrow \downarrow \right\rangle ,\left\vert \downarrow \uparrow \right\rangle ,\left\vert \downarrow \downarrow \right\rangle \right\} $. For instance,  at the state $ \left\vert \uparrow \downarrow \right\rangle =\left\vert \uparrow \right\rangle \otimes \left\vert \downarrow \right\rangle $, the label $ \uparrow $ means the spin up with quantum number $m_{\uparrow }=+\frac{1}{2}$, and $\downarrow $ means the spin down with quantum number $m_{\downarrow }=- \frac{1}{2}$. Each spin angular momentum operator is defined by local operators, and they are established by the tensor products between the identity operator and the Pauli operator $\mathbf{{\sigma}}_{z}$  where ${\mathbf{I}}_{z} =\frac{1}{2}{\sigma}_{z}  \otimes {\mathbf{1}} $ and  ${\mathbf{S}}_{z} = {\mathbf{1}} \otimes \frac{1}{2}  {\sigma}_{z}  $. The secular Hamiltonian,  ${\mathcal{H}}_{0}$,  is diagonal and the eigenenergies are detailed by the set $\left\{ -\frac{\hbar \left( \omega ^{I} + \omega ^{S} \right) }{2} ,-\frac{\hbar \left( \omega ^{I}- \omega ^{S} \right) }{2} ,\frac{\hbar \left( \omega ^{I}- \omega ^{S} \right) }{2} , \frac{\hbar \left( \omega ^{I} + \omega ^{S} \right) }{2} \right\} $ in correspondence with the basis of quantum states defined in previous paragraphs.

\begin{figure}[t]
\includegraphics[width=4.50in]{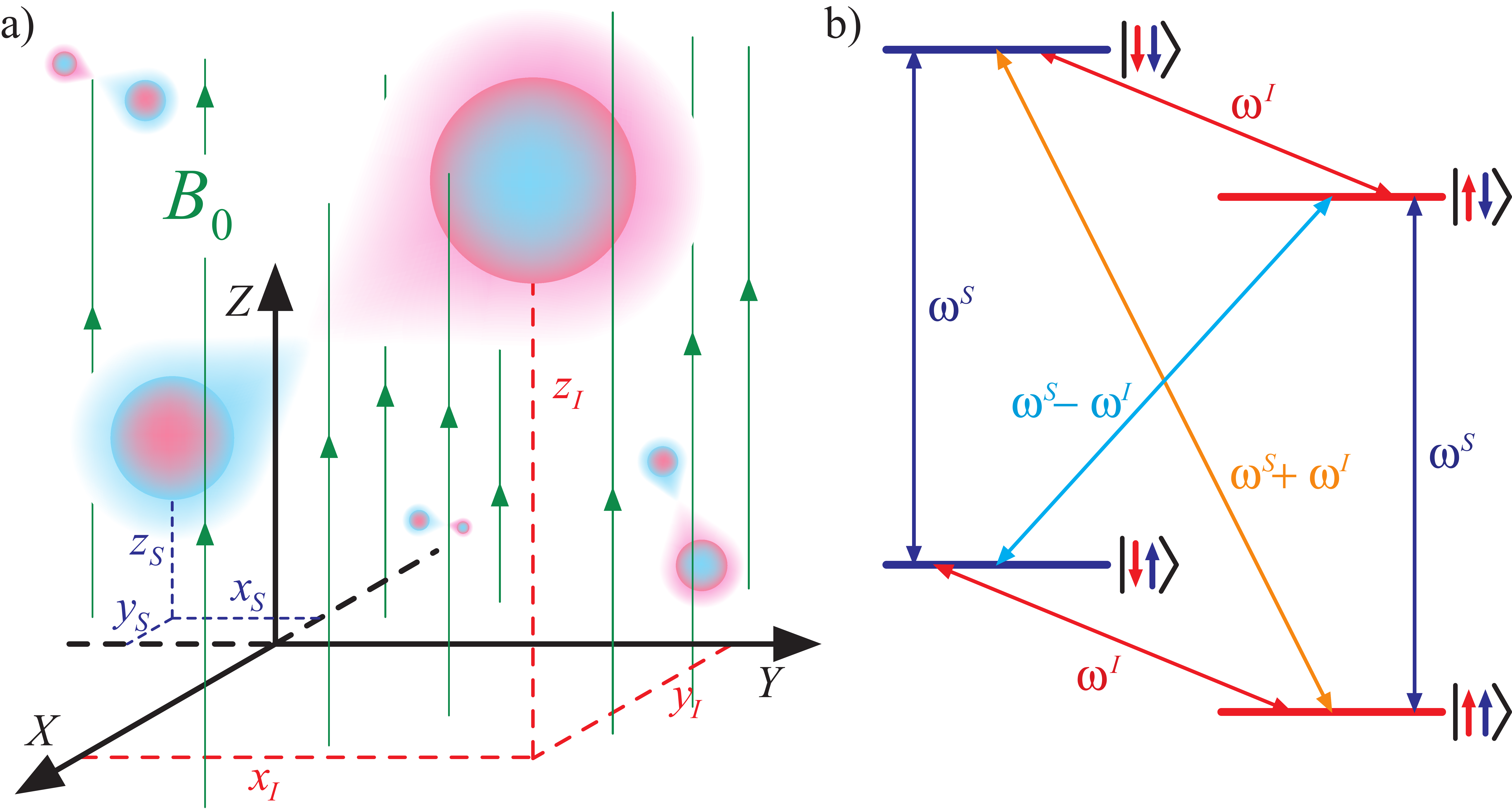}
\caption[Spectrum and Molecule]{(Color online) a) In the laboratory frame is sketched a pictorial representation of an unlike spin pair system    interacting by the dipole-dipole coupling. b) Eigenenergies levels of the secular Hamiltonian ${\mathcal{H}}_{0}$, and transition frequencies characterizing  zero ($\omega^{S}-\omega^{I} $), first ($\omega^{S} $, $\omega^{I} $), and second ($\omega^{S}+\omega^{I} $) order coherences.} \label{fig:SistemaDipolar}
\end{figure}

The dipole-dipole coupling interaction establishes an interaction between a spin particle with magnetic moment $\hbar\gamma_{I}{\mathbf{I}}_{z}$  and another neighbor spin particle with magnetic moment  $\hbar\gamma_{S}{\mathbf{S}}_{z}$. Therefore, the energy due to the interaction is quantified  from the  classical electromagnetic notation to its quantum counterpart as denoted by \cite{abragam1994Book}
\begin{equation}
{\mathcal{H}}_{D}=\frac{\hbar ^{2}\gamma ^{I}\gamma ^{S}}{r^{3}} \left( {\mathbf{I}}\cdot {\mathbf{S}}-\frac{3\left( {\mathbf{I}} \cdot \mathbf{r}\right) \left( {\mathbf{S}}\cdot \mathbf{r}\right) }{ r^{2}}\right) \text{,} \label{HamiltonianoDipolar}
\end{equation}
where $\mathbf{r}=\left( x,y,z\right) = \left( x_{I}-x_{S},y_{I}-y_{S},z_{I}-z_{S}\right)  $ means the vector of the relative spatial position between both nuclear spins and $r$ its length, a pictorial representation is sketched in Fig. \ref{fig:SistemaDipolar}a.  A convenient  notation of the dipole-dipole interaction energy can be introduced in terms of a second rank operators ${\mathbf{A}}^{\left( m\right) }$, using the ladder angular momentum operators
\begin{subequations}
\label{eq:DipolarTensorOperatorRank2}
\begin{eqnarray}
{\mathbf{A}}^{\left( -2\right) } &=&{\mathbf{I}}_{-1}%
{\mathbf{S}}_{-1}\text{,}
\label{eq:DipolarTensorOperatorRank2Menos02} \\
{\mathbf{A}}^{\left( -1\right) } &=&{\mathbf{I}}_{-1}%
{\mathbf{S}}_{0}+{\mathbf{I}}_{0}{\mathbf{S%
}}_{-1}\text{,}  \label{eq:DipolarTensorOperatorRank2Menos01} \\
{\mathbf{A}}^{\left( 0\right) } &=&\frac{2}{\sqrt{6}}\left( 3%
{\mathbf{I}}_{0}{\mathbf{S}}_{0} - {\mathbf{I%
}}\cdot  {\mathbf{S}}\right) \text{,}
\label{eq:DipolarTensorOperatorRank2Nulo00} \\
 {\mathbf{A}}^{\left( +1\right) } &=&-\left(  {\mathbf{I}}_{+1}%
 {\mathbf{S}}_{0} + {\mathbf{I}}_{0} {\mathbf{S%
}}_{+1} \right) \text{,}  \label{eq:DipolarTensorOperatorRank2Mas01} \\
{\mathbf{A}}^{\left( +2\right) } &=&{\mathbf{I}}_{+1}%
{\mathbf{S}}_{+1}\text{,} \label{eq:DipolarTensorOperatorRank2Mas02}
\end{eqnarray}
\end{subequations}
where the set of ladder angular momentum operators obeys the notation of the Eq. (4.13) in Ref. \cite{mcconnell1987Book}, and the elements of the set can be represented in terms of the canonical angular momentum operators ${\mathbf{S}}_{\pm 1} = {\mathbf{S}}_{x} \pm i {\mathbf{S}}_{y}$,  ${\mathbf{I}}_{\pm 1} = {\mathbf{I}}_{x} \pm i {\mathbf{I}}_{y}$,   ${\mathbf{S}}_{0} = {\mathbf{S}}_{z} $,   and  ${\mathbf{I}}_{0} = {\mathbf{I}}_{z} $. Also, each of these  second-rank operators, ${\mathbf{A}}^{\left( m\right) }$ is multiplied by spherical harmonic functions of rank 2 denoted by $Y_{l}^{m}\left(\theta,\phi\right)$ with $l=2$ and $m=\pm2,\pm1,0$; where $\theta$ is the angle between the $Z$-axis and the vector $\mathbf{r}$ (see Fig. \ref{fig:SistemaDipolar}a); $\phi$ is the angle between the $X$-axis and the projection of the vector $\mathbf{r}$ onto the $XY$-plane. Therefore, the dipole-dipole Hamiltonian operator is rewritten by
\begin{equation}
{\mathcal{H}}_{D}=-\frac{\hbar ^{2}\gamma ^{I}\gamma ^{S}}{r^{3}} \sqrt{\frac{6\pi }{5}}\sum_{m=-2}^{+2}\left( -1\right) ^{m}\left( Y_{2}^{-m}\left(\theta,\phi\right)\right) \left( {\mathbf{A}}^{\left( m\right) }\right) \text{,} \label{HamiltonianoDipolarExplicito}
\end{equation}
this Hamiltonian matches with  Eq.  (4.17) of Ref. \cite{mcconnell1987Book}. Considering these both energy contributions, the total Hamiltonian of the two-level spin systems pair is represented in the laboratory frame description by
\begin{equation}
{\mathcal{H}}={\mathcal{H}}_{0}+{\mathcal{H}}_{D}\text{.} \label{HamiltonianoTotal}
\end{equation}

Following the standard procedures to discuss the relaxation dynamics of nuclear spins, some assumptions must be introduced to help its description. One of them is an explicit definition of the density matrix in its matrix notation   
\begin{equation}
{\mathbf{\varrho }}=\left[ 
\begin{array}{cccc}
\varrho _{1,1} & \varrho _{1,2} & \varrho _{1,3} & \varrho _{1,4} \\ 
\varrho _{2,1} & \varrho _{2,2} & \varrho _{2,3} & \varrho _{2,4} \\ 
\varrho _{3,1} & \varrho _{3,2} & \varrho _{3,3} & \varrho _{3,4} \\ 
\varrho _{4,1} & \varrho _{4,2} & \varrho _{4,3} & \varrho _{4,4}%
\end{array}%
\right] \text{.}  \label{MatricialRepresentationDensityOperator}
\end{equation}
The explicit density operator indexing illustrates each element's position on the matrix. Considering the indexing and the measurable observables along the z-axis (or the longitudinal magnetization), we know that any physical measurement will depend only on diagonal elements, i.e., a four-dimensional dependence. Also, suppose the Redfield theory was applied, assuming the mean values of some measurable observables and other multiplications between pairs of operators. In that case, the main results will preserve the previous findings (see, for example, Eq. (3.19-3.21) of Ref. \cite{kowalewski2018Book} or Eq. (4.78-4.80) of Ref. \cite{mcconnell1987Book}). Alternatively, suppose the Redfield theory is applied to study the same physical system under the same assumptions but to solve for each density matrix element. In that case, it will generate a linear system of dimension six. This new value and its consequences will be explored in this study. The dimensional value 6 matches the meaning of the number of density matrix elements of zero-order coherence. This criterion was applied to spin systems with $I > 1/2$ spin value \cite{consuelo-leal2023JMR} or identical many-bodies of two-level particles \cite{seminara1990}.

The other  assumption is related to establishing an equation of motion to predict the dynamics of an open quantum system. At the regime of high temperature, the dynamics of the spin system interacting weakly with an environment can be appropriately described  by Redfield theory. The standard description is developed using the second-order approximation of the Liouville-von Neumann equation in the Interaction picture \cite{redfield1957}. 

An equation of motion for the density operator in terms of the double commutator operator with the Hamiltonian of perturbation, on the one hand, and a time derivative operation, on the other, is generated and denoted by
\begin{eqnarray}
\frac{d\widetilde{\varrho} \left( t\right) }{dt} &=&  i\left[ \widetilde{\varrho} \left( t\right) , \widetilde{\mathcal{H}}_{D}\left( t\right) \right] -\int\limits_{0}^{\infty}\left[ \left[ \widetilde{\varrho} \left( t\right) ,\widetilde{\mathcal{H}}_{D}\left( t^{\prime }\right) \right] ,\widetilde{\mathcal{H}}_{D}\left( t\right) \right] dt^{\prime }\text{,} \label{EquationMotionByRedfieldTheory}
\end{eqnarray}
where $\widetilde{\mathcal{O}} \left( t\right)$ means  any operator in the Interaction picture. The first term of the equation of motion describes the dynamics of an isolated system. The second term describes the relaxation dynamics of the system under the effects of an environment. These effects can be encoded in the double commutator of the Eq. (\ref{EquationMotionByRedfieldTheory}), which depends on the dipole-dipole Hamiltonian of Eq. (\ref{HamiltonianoDipolarExplicito})  denoted in the interaction picture $\widetilde{\mathcal{H}}_{D} \left( t^{\prime}\right)$ and  $\widetilde{\mathcal{H}}_{D} \left( t\right)$. In principle, the fluctuations in the relative orientation of a vector $\mathbf{r} \left( t\right)$, used to establish the dipole-dipole Hamiltonian at different time $t$ and $t^{\prime}$, describes the appropriate sources of the relaxation phenomena. It means that there is a time dependence of the dipole-dipole coupling energy at time $t^{\prime}$, and this time dependence can be denoted using angular parameters of the spherical harmonic functions $Y_{2}^{-m^{\prime}}\left(\theta \left( t^{\prime}\right),\phi \left( t^{\prime}\right)\right)$. Similarly, after a time $t $, the time dependence of the dipole-dipole coupling energy is denoted by $Y_{2}^{-m}\left(\theta \left( t\right),\phi \left( t\right)\right)$. We want to know if both events are correlated or uncorrelated, then we must compute the integral $\int\limits_{0}^{\infty} \cdots dt^{\prime}$. To do the computation, it is introduced a time parameter named correlation time defined by $\tau = t-t^{\prime}$ such that from the computation of the integral, the spectral density functions were defined in terms of characteristic frequencies and the correlation times, $J \equiv J \left( \omega  ,\tau  \right)$. 

Therefore,  performing a sequence of standard algebraic procedures, the relaxation superoperator  representation was denoted by four terms, which were described appropriately by the multiplication between  pairs of second-rank operators elements  ${\mathbf{A}}^{\left( m\right) }$   ${\mathbf{A}}^{\left( m^{\prime}\right) }$ and  spectral density functions $J\left( \omega,\tau \right)$. 
Consequently, the equation of motion for each density matrix element is analyzed by performing some principles of algebra to describe the matrix representation of operators, properties of commutation rules between angular momentum operators, and introducing a concise notation as described in Eq. (3.58) of Ref. \cite{mcconnell1987Book}. Considering those three formal mathematical procedures, the time evolution of each element  $\varrho_{\alpha, \beta}$  is denoted by  the following differential equation 
\begin{equation}
\frac{\partial {\varrho }_{\alpha, \beta}\left( t\right) }{\partial t }= i\left[ {\varrho} \left( t\right) , {\mathcal{H}}_{D}\left( t\right) \right]_{\alpha, \beta} + \sum\limits_{ \alpha ^{\prime } \beta ^{\prime }}\mathcal{R}_{\alpha \beta}^{\alpha ^{\prime } \beta ^{\prime }}{\varrho }_{\alpha ^{\prime }, \beta ^{\prime }}\left( t\right),  \label{EquacaoRedfield}
\end{equation}
where $\left[ {\varrho} \left( t\right) , {\mathcal{H}}_{D}\left( t\right) \right]_{\alpha, \beta} \equiv \langle \alpha \vert \left[ {\varrho} \left( t\right) , {\mathcal{H}}_{D}\left( t\right) \right] \vert \beta \rangle$ means a projection of the commutator operator under the elements $ \vert \alpha \rangle$ and $\vert \beta \rangle$ of the basis (this term will be neglected in this analysis because it does not represent any relaxation effect),   and $\mathcal{R}_{\alpha\beta}^{ \alpha ^{\prime } \beta ^{\prime }}$  represents the  transformation  element $\alpha \beta \alpha ^{\prime }\beta ^{\prime }$ of the relaxation superoperator \cite{redfield1957,abragam1994Book}. Therefore, in the present description of the  Redfield equations, we denote them as follows
\begin{eqnarray}
\frac{1}{\mathcal{C}}\frac{d\varrho _{\alpha ,\beta }\left( t\right) }{dt} &=&+\sum_{\alpha ^{\prime   },\beta ^{\prime   }} J\left( \omega _{\alpha } -\omega _{\alpha ^{\prime   }} \right) \left( \sum_{p=-2}^{+2}\left( -1\right) ^{p}\left( {\mathbf{A}}_{\alpha ,\alpha ^{\prime  }}^{\left( -p\right) }\right) \left( { \mathbf{A}}_{\beta ^{\prime  },\beta }^{\left( p\right) }\right) \right)   \varrho _{\alpha ^{\prime  },\beta ^{\prime  }}\left( t\right) \notag \\
&&+\sum_{\alpha ^{\prime  },\beta ^{\prime  }} J\left( \omega _{\beta ^{\prime  }} -\omega _{\beta } \right) \left( \sum_{p=-2}^{+2}\left( -1\right) ^{p}\left( {\mathbf{A}}_{\alpha ,\alpha ^{\prime  }}^{\left( -p\right) }\right) \left( { \mathbf{A}}_{\beta ^{\prime  },\beta }^{\left( p\right) }\right) \right) \varrho _{\alpha ^{\prime  },\beta ^{\prime  }}\left( t\right) \notag \\
&&-\sum_{\alpha ^{\prime   },\beta ^{\prime   }} J\left( \omega _{\beta ^{\prime   }} -\omega _{\alpha ^{\prime }} \right) \left( \sum_{p=-2}^{+2}\left( -1\right) ^{p}\left( { \mathbf{A}}_{\alpha ^{\prime   },\beta ^{\prime   }}^{\left( -p\right) }\right) \left( {\mathbf{A}}_{\beta ^{\prime },\beta }^{\left( p\right) }\right) \right) \varrho _{\alpha ,\alpha ^{\prime   }}\left( t\right) \notag \\
&&-\sum_{\alpha ^{\prime   },\beta ^{\prime   }} J\left( \omega _{\alpha ^{\prime   }} -\omega _{\beta ^{\prime }} \right) \left( \sum_{p=-2}^{+2}\left( -1\right) ^{p}\left( { \mathbf{A}}_{\alpha ,\alpha ^{\prime   }}^{\left( -p\right) }\right) \left( {\mathbf{A}}_{\alpha ^{\prime   },\beta }^{\left( p\right) }\right) \right) \varrho _{\beta ^{\prime   },\beta }\left( t\right) \text{,} \quad \quad
\label{EquationOfRelaxationDipolarHamiltonian04TermosComOperadorG}
\end{eqnarray}
where operators ${\mathbf{A}}^{\left( p\right) }$ were defined in Eq.  (\ref{eq:DipolarTensorOperatorRank2}) (or see Eq. (4.18) of Ref. \cite{mcconnell1987Book}) and $\mathcal{C}$ is a constant parameter that depends on nuclear magnetic moments of nuclei, the length $r$ between  nuclear magnetic moments, and the reduced Planck's constant.   The constant parameter $\mathcal{C}$ is denoted by Eq. (5) of Ref. \cite{boros2018}, or Eq. (42) of Ref. \cite{solomon1955}, 
\begin{equation}
\mathcal{C} =  \left(   \frac{\mu_{0}}{4\pi} \frac{\gamma_{I}\gamma_{S} \hbar }{r^{3}}  \right) ^{2} \left(   \frac{1}{2} \left(  \frac{1}{2} +1\right) \right)     \  \propto  \   \left(     \frac{\gamma_{I}\gamma_{S}\hbar }{r^{3}}  \right) ^{2}    \text{.}
\label{ConstantProportionalityGeneral}
\end{equation}
This representation of the Redfield equation allows a direct application of algebraic methods and appropriate index counting, both were used to generate linear systems of differential equations. Eq. (\ref{EquationOfRelaxationDipolarHamiltonian04TermosComOperadorG}) and a convenient  choice  of the dummy subscripts to label any density matrix element denoted by $\alpha ,\beta=1,2,3,4$ were used to establish the linear system of differential equations and to analyze the spin system dynamics. Also, an advantage of Eq. (\ref{EquationOfRelaxationDipolarHamiltonian04TermosComOperadorG}) is the ability to distinguish the algebraic rules and physical meaning. From the algebraic rules, the linear superposition performed by the evaluation of the sum operation subscripts $\alpha^{\prime}$,  $\beta^{\prime}$, and $p$ is reduced thanks to some null elements of the second-rank operators (see Appendix). Performing the physical information of the relaxation procedure, which was managed by the constant coefficient $\mathcal{C}$ and the spectral density functions $J\left( \omega , \tau \right)$, the parity property of the spectral density function improves some simplification of other terms. The mutual action of both abilities will be highlighted in the applications.

\subsection{Study of an unlike spin pair system}
\label{sec:TwoNucleiSpinSystemTypeAX}

The discussion of any unlike spin pair system is compatible with any pair of the heteronuclear spin system, and many classifications have been made along with new developments and established their characteristics \cite{corio1967Book}. From  a theoretical point of view, relaxation process of any pair of the heteronuclear spin system is described more appropriately, defining sequential labeling of an eigenstate basis,  such as \ $\left\{ \left\vert \uparrow \uparrow \right\rangle ,\left\vert \uparrow \downarrow \right\rangle ,\left\vert \downarrow \uparrow \right\rangle ,\left\vert \downarrow \downarrow \right\rangle \right\} \equiv \left\{ \left\vert 1\right\rangle ,\left\vert 2\right\rangle  ,\left\vert 3\right\rangle ,\left\vert 4\right\rangle \right\} $ (see Sec. 2.1 of Ref. \cite{kruk2016Book}). Likewise, their respective eigenenergies must be computed from the characteristic equations of the secular Hamiltonian operator ${\mathcal{H}}_{0}$. They  are summarized as follows 
\begin{subequations}
\label{EigenEnergyDipolarTypeAX}
\begin{eqnarray}
{\mathcal{H}}_{0}\left\vert 1\right\rangle &=&\hbar \omega _{1}\left\vert 1\right\rangle =\hbar \left( -\frac{ 
\omega ^{I}+\omega ^{S}  }{2}\right) \left\vert 1\right\rangle \text{,}
\label{EigenEnergyDipolarTypeAXQS1} \\
{\mathcal{H}}_{0}\left\vert 2\right\rangle &=&\hbar \omega _{2}\left\vert 2\right\rangle =\hbar \left( -\frac{ 
\omega ^{I}-\omega ^{S}  }{2}\right) \left\vert 2\right\rangle \text{,}
\label{EigenEnergyDipolarTypeAXQS2} \\
{\mathcal{H}}_{0}\left\vert 3\right\rangle &=&\hbar \omega _{3}\left\vert 3\right\rangle =\hbar \left( +\frac{ 
\omega ^{I}-\omega ^{S} }{2}\right) \left\vert 3\right\rangle \text{,}
\label{EigenEnergyDipolarTypeAXQS3} \\
{\mathcal{H}}_{0}\left\vert 4\right\rangle &=&\hbar \omega _{4}\left\vert 4\right\rangle =\hbar \left( +\frac{ 
\omega ^{I}+\omega ^{S}  }{2}\right) \left\vert 4\right\rangle \text{.}
\label{EigenEnergyDipolarTypeAXQS4}
\end{eqnarray}%
\end{subequations}
Therefore, applying conveniently  Eq. (\ref{EquationOfRelaxationDipolarHamiltonian04TermosComOperadorG}-\ref{EigenEnergyDipolarTypeAX}) and an appropriate  choice of the dummy indexes,  let us start to compute the three sets of linear differential equations for zero, first, and second-order density matrix elements. To perform a concise notation of the spectral density functions is introduced the following ones  $J\left( 0\right) =J_{0}$, $J\left( \omega ^{S}\right) =J_{S}$, $J\left( \omega ^{I}\right) =J_{I}$, $J\left( \omega ^{S}+\omega ^{I}\right) =J_{+}$, and $J\left( \omega ^{S}-\omega ^{I}\right) =J_{-}$.

The first set is devoted to computing the solution of the zero-order elements, and in this set belongs six equations with six parameters as resumed in the linear system
\begin{equation}
\frac{1}{\mathcal{C}}
\left[ 
\begin{array}{c}
\frac{d\delta\rho _{1,1}\left( t\right) }{dt} \\ 
\frac{d\delta\rho _{2,2}\left( t\right) }{dt} \\ 
\frac{d  \varrho _{2,3}\left( t\right)  }{dt} \\ 
\frac{d  \varrho _{3,2}\left( t\right)  }{dt} \\ 
\frac{d\delta\rho _{3,3}\left( t\right) }{dt} \\ 
\frac{d\delta\rho _{4,4}\left( t\right) }{dt}%
\end{array}%
\right] =  {\boldsymbol{\mathcal{J}}}^{\left( 0\right)}  \left[ 
\begin{array}{c}
\delta\rho _{1,1}\left( t\right) \\ 
\delta\rho _{2,2}\left( t\right) \\ 
 \varrho _{2,3}\left( t\right)  \\ 
\varrho _{3,2}\left( t\right)  \\ 
\delta\rho _{3,3}\left( t\right) \\ 
\delta\rho _{4,4}\left( t\right)
\end{array}
\right] \text{,}
\end{equation}
where $\boldsymbol{\mathcal{J}}^{\left( 0\right)}$ denotes the Redfield superoperator of zero order and it is represented using the matrix notation by

\begin{eqnarray}
\boldsymbol{\mathcal{J}}^{(0)} &=&\left[ 
\begin{array}{cccc}
-\frac{J_{S}+J_{I}+4J_{+}}{2} & +\frac{J_{S}}{2} & +\frac{J_{S}+J_{I}}{4} &+\frac{J_{S}+J_{I}}{4}
\\ 
+\frac{J_{S}}{2} & -\frac{3J_{S}+3J_{I}+2J_{-}}{6} & -\frac{J_{S}+J_{I}}{4}
& -\frac{J_{S}+J_{I}}{4} \\ 
+\frac{J_{S}+J_{I}}{4} & -\frac{J_{S}+J_{I}}{4} & -\frac{3J_{S}+3J_{I}+2J_{-}}{6} & +\frac{J_{-}}{3} \\ 
+\frac{J_{S}+J_{I}}{4} & -\frac{J_{S}+J_{I}}{4} & +\frac{J_{-}}{3} & -\frac{3J_{S}+3J_{I}+2J_{-}}{6} \\ 
+\frac{J_{I}}{2} & +\frac{J_{-}}{3} & -\frac{J_{S}+J_{I}}{4} & -\frac{J_{S}+J_{I}}{4}\\
+\left( 2J_{+}\right) & +\frac{J_{I}}{2} & +\frac{J_{S}+J_{I}}{4} & +\frac{
J_{S}+J_{I}}{4}
\end{array}
\right.  \nonumber \\ 
&&\left. 
\begin{array}{cc}
 +\frac{J_{I}}{2}  & +2J_{+} \\ 
+\frac{J_{-}}{3} & +\frac{J_{I}}{2}\\ 
-\frac{J_{S}+J_{I}}{4} & +\frac{J_{S}+J_{I}}{4}  \\ 
-\frac{3J_{S}+3J_{I}+2J_{-}}{6} & +\frac{J_{S}}{2}\\ 
+\frac{J_{S}}{2} & -\frac{J_{S}+J_{I}+4J_{+}}{2}
\end{array}%
\right] \label{eq:SuperOperator0thOrder} \text{,}
\end{eqnarray}
where only the diagonal elements of the density matrix obey  $\delta\rho _{i,i}\left( t\right) =\varrho _{i,i}\left( t\right) -\varrho _{i,i}^{\text{eq}}$ with dependence on the steady state of thermal equilibrium as denoted by $\varrho _{i,i}^{\text{eq}}$ (known in experimental quantum information studies \cite{consuelo-leal2023JMR,dieguez2022,auccaise2011B,consuelo-leal2019}). The definition of these differences does not introduce any mathematical or formal consequence on the solution. It is a practical procedure to identify a concise notation for a boundary condition of the linear system of differential equations. An advantage of this assumption is the possibility of choosing the steady state, i.e., a pure quantum state or a mixed quantum state, both of them are suitable with the secular Hamiltonian ${\mathcal{H}}_{0}$.

The second set of linear differential equations was computed, and  identified four equations that correspond to the first-order density matrix elements as denoted
\begin{equation}
\frac{1}{\mathcal{C}}
\left[ 
\begin{array}{c}
\frac{d\varrho _{1,2}\left( t\right) }{dt} \\ 
\frac{d\varrho _{1,3}\left( t\right) }{dt} \\ 
\frac{d\varrho _{2,4}\left( t\right) }{dt} \\ 
\frac{d\varrho _{3,4}\left( t\right) }{dt}%
\end{array}%
\right] =   \boldsymbol{\mathcal{J}}^{\left( 1\right)}  \left[ 
\begin{array}{c}
\varrho _{1,2}\left( t\right) \\ 
\varrho _{1,3}\left( t\right) \\ 
\varrho _{2,4}\left( t\right) \\ 
\varrho _{3,4}\left( t\right)%
\end{array}
\right] \text{,}
\end{equation}
where $\boldsymbol{\mathcal{J}}^{\left( 1\right)}$ denotes the Redfield superoperator of first-order and it is represented using the matrix notation by
\begin{eqnarray}
\boldsymbol{\mathcal{J}}^{(1)} &=&\left[ 
\begin{array}{cc}
-  \frac{4J_{0}+3J_{S}+3J_{I}+J_{-}+6J_{+}}{6} & - \frac{4J_{0}+3J_{S}+3J_{I}+4J_{-}}{12} \\
- \frac{4J_{0}+3J_{S}+3J_{I}+4J_{-}}{12} & -\frac{4J_{0}+3J_{S}+3J_{I}+J_{-}+6J_{+}}{6}\\
- \frac{J_{S}+J_{I}}{4} & -\frac{J_{S}}{2}\\
-\frac{J_{I}}{2} & - \frac{J_{S}+J_{I}}{4}
\end{array}
\right.  \nonumber \\ 
&&\left. 
\begin{array}{cc}
- \frac{J_{S}+J_{I}}{4} & -\frac{J_{I}}{2} \\
-\frac{J_{S}}{2} & - \frac{J_{S}+J_{I}}{4} \\
-\frac{4J_{0}+3J_{S}+3J_{I}+J_{-}+6J_{+}}{6} & - \frac{4J_{0}+3J_{S}+3J_{I}+4J_{-}}{12}\\
- \frac{4J_{0}+3J_{S}+3J_{I}+4J_{-}}{12} & -\frac{4J_{0}+3J_{S}+3J_{I}+J_{-}+6J_{+}}{6}
\end{array}%
\right] \label{eq:SuperOperator1stOrder}\text{.}
\end{eqnarray}

The third set of linear system of differential equations was computed such that it was found one  which represents the second-order  density  matrix element as denoted
\begin{equation}
\frac{1}{\mathcal{C}}
\frac{d\varrho _{1,4}\left( t\right) }{dt}=  \boldsymbol{\mathcal{J}}^{\left( 2\right)}  \varrho _{1,4}\left( t\right) = -  \frac{J_{S}+J_{I}+4J_{+}}{2}  \varrho _{1,4}\left( t\right)  \label{eq:SuperOperator2ndOrder} \text{,}
\end{equation}
where $\boldsymbol{\mathcal{J}}^{\left( 2\right)}$ denotes the Redfield superoperator of second-order, represented by only one matrix element.

The solutions of each set of linear differential equations were computed using algebraic matrix properties and standard procedures to solve ordinary differential equations. The solution is resumed with the computation of the eigenvalues of Redfield superoperators at each coherence order $\iota_{\beta}^{\left( \alpha\right) }$  detailed in Tab. \ref{tab:RelaxationRatesAX}.

In the case of the zero-order density  matrix elements, the solution corresponded with an appropriate superposition of exponential functions and it was represented in terms of the relaxation rate constants denoted by $ R _{\beta}^{\left( 0\right) }=\mathcal{C} \ \iota_{\beta}^{\left( 0 \right) }  $, where $\iota_{\beta}^{\left( 0 \right) }$ represents any eigenvalue of the superoperator $\boldsymbol{\mathcal{J}}^{\left( 0\right)}$, and those can be summarized in Tab. \ref{tab:RelaxationRatesAX}. Therefore, the solutions are written as follows
\begin{subequations}
\label{SpinSystemAXZeroOrder}
\begin{eqnarray}
\delta\rho _{1,1}( t)  &=&+\Xi_{b}^{\left( 0\right) }\left( t\right) \varrho _{B}\left( t_{0}\right) + \varrho _{E}\left( t_{0}\right) -\Xi_{f}^{\left( 0\right) }\left( t\right) \varrho _{H}\left( t_{0}\right) -\Xi_{g}^{\left( 0\right) }\left( t\right)  \varrho _{J}\left(t_{0}\right) \text{,} \quad \quad \quad  \label{SpinSystemAXZeroOrderRho11}\\
\delta\rho _{2,2}( t)  &=&-\Xi_{b}^{\left( 0\right) }\left( t\right) \varrho _{B}\left( t_{0}\right) + \varrho _{C}\left( t_{0}\right) +\Xi_{f}^{\left( 0\right) }\left( t\right) \varrho _{F}\left( t_{0}\right) +\Xi_{g}^{\left( 0\right) }\left( t\right) \varrho _{G}\left(t_{0}\right) \text{,}  \label{SpinSystemAXZeroOrderRho22} \\
\varrho _{2,3}\left( t\right)  &=&-\Xi_{a}^{\left( 0\right) }\left( t\right) \varrho _{A}\left( t_{0}\right) -\Xi_{b}^{\left( 0\right) }\left( t\right) \varrho _{B}\left( t_{0}\right) -  \varrho _{C}\left( t_{0}\right) +  \varrho _{E}\left( t_{0}\right) \text{,}  \label{SpinSystemAXZeroOrderRho23} \\
\varrho _{3,2}\left( t\right)  &=&+\Xi_{a}^{\left( 0\right) }\left( t\right) \varrho _{A}\left( t_{0}\right) -\Xi_{b}^{\left( 0\right) }\left( t\right) \varrho _{B}\left( t_{0}\right) -  \varrho _{C}\left( t_{0}\right) +  \varrho _{E}\left( t_{0}\right) \text{,}  \label{SpinSystemAXZeroOrderRho32} \\
\delta\rho _{3,3}\left( t\right)  &=&-\Xi_{b}^{\left( 0\right) }\left( t\right) \varrho _{B}\left( t_{0}\right) +  \varrho _{C}\left( t_{0}\right) -\Xi_{f}^{\left( 0\right) }\left( t\right) \varrho _{F}\left( t_{0}\right) -\Xi_{g}^{\left( 0\right) }\left( t\right) \varrho _{G}\left( t_{0}\right) \text{,}  \label{SpinSystemAXZeroOrderRho33} \\
\delta\rho _{4,4}\left( t\right)  &=&+\Xi_{b}^{\left( 0\right) }\left( t\right) \varrho _{B}\left( t_{0}\right) +  \varrho _{E}\left( t_{0}\right) +\Xi_{f}^{\left( 0\right) }\left( t\right) \varrho _{H}\left( t_{0}\right) +\Xi_{g}^{\left( 0\right) }\left( t\right) \varrho _{J}\left(t_{0}\right) \text{,} \label{SpinSystemAXZeroOrderRho44}
\end{eqnarray}
\end{subequations}
where $\Xi_{\xi}^{\left( 0\right) }\left( t\right)$ means the exponential function depending on relaxation rates of zero order as denoted by $ \exp \left[ -R_{\xi}^{\left( 0\right) }\left( t-t_{0}\right) \right]$ with the subscript $\xi=a,b,c,f,g$, and only the diagonal elements of the density matrix obey  $\delta\rho _{i,i}\left( t\right) =\varrho _{i,i}\left( t\right) -\varrho _{i,i}^{\text{eq}}$. From the principles of algebra and their solutions found, each zero-order density matrix element must be represented by a superposition of six exponential functions. However, considering an appropriate linear combination of the elements and their coefficients, each density matrix element expression is represented by the superposition of only four exponential ones multiplying conveniently by their coefficients which are denoted as follows
\begin{subequations}
\label{SpinSystemAXZeroOrderCoefficients}
\begin{eqnarray}
2\varrho _{A}\left( t_{0}\right)  &=& \varrho _{3,2}\left( t_{0}\right) -\varrho _{2,3}\left( t_{0}\right)   \text{,} \label{SpinSystemAXZeroOrderCoefficientsA} \\
6\varrho _{B}\left( t_{0}\right)  &=& \delta \rho _{1,1}\left( t_{0}\right) -\delta \rho _{2,2}\left( t_{0}\right) -\varrho _{2,3}\left( t_{0}\right) -\varrho _{3,2}\left( t_{0}\right) -\delta \rho _{3,3}\left( t_{0}\right) \nonumber \\
&& +\delta \rho _{4,4}\left( t_{0}\right) \text{,}  \label{SpinSystemAXZeroOrderCoefficientsB} \\
6\varrho _{C}\left( t_{0}\right)  &=& \delta \rho _{1,1}\left( t_{0}\right) +2 \delta \rho _{2,2}\left( t_{0}\right)  -\varrho_{2,3}\left( t_{0}\right) -\varrho _{3,2}\left( t_{0}\right) +2\delta \rho_{3,3}\left( t_{0}\right) \nonumber \\
&& +\delta \rho _{4,4}\left( t_{0}\right) \text{,}  \label{SpinSystemAXZeroOrderCoefficientsC} \\
6\varrho _{E}\left( t_{0}\right)  &=& 2\delta \rho _{1,1}\left( t_{0}\right) + \delta \rho _{2,2}\left( t_{0}\right)   +\varrho_{2,3}\left( t_{0}\right) +\varrho _{3,2}\left( t_{0}\right)+\delta \rho_{3,3}\left( t_{0}\right) \nonumber \\
&&  +2\delta \rho _{4,4}\left( t_{0}\right)  \text{,} \label{SpinSystemAXZeroOrderCoefficientsE} \\
4\varrho _{F}\left( t_{0}\right)  &=&   -\frac{\left( A+B\right)\left( A-B\right)}{\left( AD-BC\right)}   \delta \rho _{1,1}\left( t_{0}\right)   -\frac{\left( A+B\right)\left( C-D\right)}{\left( AD-BC\right)}   \delta \rho _{2,2}\left( t_{0}\right) \nonumber \\
&&  +\frac{\left( A+B\right)\left( C-D\right)}{\left( AD-BC\right)}   \delta \rho _{3,3}\left( t_{0}\right)   +\frac{\left( A+B\right)\left( A-B\right)}{\left( AD-BC\right)}   \delta \rho _{4,4}\left( t_{0}\right)    \text{,}  \label{SpinSystemAXZeroOrderCoefficientsF} \\
4\varrho _{G}\left( t_{0}\right)  &=&\frac{  \left( A-B\right) \left( A+B\right)  }{\left( AD-BC\right) }   \delta \rho _{1,1}\left( t_{0}\right)   +\frac{\left( A-B\right) \left(C+D\right)   }{\left( AD-BC\right) }  \delta \rho _{2,2}\left( t_{0}\right) \nonumber \\
&&   -\frac{\left( A-B\right) \left(C+D\right)  }{\left( AD-BC\right) }      \delta \rho _{3,3}\left( t_{0}\right)   -\frac{\left( A-B\right) \left(A+B\right)   }{\left( AD-BC\right) }     \delta \rho _{4,4}\left( t_{0}\right) \text{,}  \label{SpinSystemAXZeroOrderCoefficientsG} \\
4\varrho _{H}\left( t_{0}\right)  &=&-\frac{\left( C+D\right) \left(A-B\right) }{\left( AD-BC\right) }   \delta \rho _{1,1}\left( t_{0}\right)   -\frac{\left( C+D\right) \left(C-D\right) }{\left( AD-BC\right) }   \delta \rho _{2,2}\left( t_{0}\right) \nonumber \\
&&  +\frac{\left( C+D\right) \left(C-D\right) }{\left( AD-BC\right) }   \delta \rho _{3,3}\left( t_{0}\right)   +\frac{\left( C+D\right) \left(A-B\right) }{\left( AD-BC\right) }   \delta \rho _{4,4}\left( t_{0}\right) \text{,}  \label{SpinSystemAXZeroOrderCoefficientsH} \\
4\varrho _{J}\left( t_{0}\right)  &=&\frac{\left( C-D\right) \left(A+B\right)}{\left( AD-BC\right) }   \delta \rho _{1,1}\left( t_{0}\right)   +\frac{\left( C-D\right) \left(C+D\right)}{\left( AD-BC\right) }   \delta \rho _{2,2}\left( t_{0}\right) \nonumber \\
&&  -\frac{\left( C-D\right) \left( C+D\right)}{\left( AD-BC\right) }   \delta \rho _{3,3}\left( t_{0}\right)   -\frac{\left( C-D\right) \left(A+B\right)}{\left( AD-BC\right) }   \delta \rho _{4,4}\left( t_{0}\right) \text{,} \qquad \qquad \label{SpinSystemAXZeroOrderCoefficientsJ}
\end{eqnarray}
\end{subequations}
where the $A$, $B$, $C$, and $D$ parameters are defined in terms of spectral density functions
\begin{subequations}
\label{eq:NormaTransParam}
\begin{eqnarray}
A &=& 27\left( J_{S}-J_{I}\right) \left( J_{S}+J_{I}\right) ^{3} +4\left( J_{S}-J_{I}\right) \left( 6J_{+}+J_{-}\right) \left( 3J_{S}J_{I} \right. \nonumber \\
&& \left. -4\left( 6J_{+}-J_{-}\right) ^{2} +24\left( J_{S}J_{I}-J_{+}J_{-}\right)\right)  \text{,}  \label{eq:NormaTransParamA} \\
B &=&54\left( J_{S}-J_{I}\right)  \left(\left( J_{S}+J_{I}\right)^{2}+2J_{S}J_{I}-\frac{8}{9}\left( 6J_{+}+J_{-}\right) ^{2}+\frac{16}{3}J_{+}J_{-}\right)  \mho \text{,}  \label{eq:NormaTransParamB} \\
C &=&27\left( J_{S}+J_{I}\right) \left( J_{S}^{3}+J_{I}^{3}\right) +18\left(
J_{S}\ +J_{I}\right) ^{3}\left( 6J_{+}-J_{-}\right)  -288\left(J_{S}^{2}+J_{I}^{2}\right) J_{+}J_{-} \nonumber \\
&& +9\left( 32J_{+}\left( 6J_{+}+J_{-}\right) +24\left( J_{S}J_{I}-16J_{+}^{2}\right) -3\left( J_{S}-J_{I}\right) ^{2}\right) \left( 16J_{+}^{2}-J_{S}J_{I}\right) \nonumber \\
&&+108J_{S}^{2}J_{I}^{2}-12\left( J_{S}-J_{I}\right) ^{2}J_{-}^{2} \text{,}  \label{eq:NormaTransParamC} \\
D &=&\left( 54\left( J_{S}+J_{I}\right) ^{3}+36\left( \left(
J_{S}+J_{I}\right) ^{2}+4J_{S}J_{I}\right) \left( 6J_{+}-J_{-}\right)
\right) \mho   \nonumber \\
&&    + 288\left(J_{S}J_{I}J_{-}-48J_{+}^{3}+3J_{S}J_{I}J_{+}\right)\mho\text{.}
\label{eq:NormaTransParamD}
\end{eqnarray}
\end{subequations}
where $\mho$ represents the radical with the addition of two squared differences of spectral density functions as defined by
\begin{equation}
\mho=\sqrt{\left(\frac{ J_{S}-J_{I}}{2}\right) ^{2}+\left(\frac{ 6J_{+}-J_{-}}{3}\right) ^{2}} \text{.}
\end{equation}

An analogous procedure is developed to compute the linear system solutions for the first-order density matrix elements. Four eigenvalues characterize this linear system. Consequently, four exponential functions multiplying each appropriate coefficient can describe each density matrix element, and these are written as follows
\begin{subequations}
\label{SpinSystemAXFirstOrder}
\begin{eqnarray}
\varrho _{1,2}\left( t\right)  &=&-\Xi _{a}^{\left( 1\right) }\varrho
_{K_{ab}}\left( t_{0}\right) -\Xi _{b}^{\left( 1\right) }\varrho
_{L_{ab}}\left( t_{0}\right) +\Xi _{f}^{\left( 1\right) }\varrho
_{K_{fg}}\left( t_{0}\right) +\Xi _{g}^{\left( 1\right) }\varrho
_{L_{fg}}\left( t_{0}\right) \text{,} \quad \quad \quad \label{SpinSystemAXFirstOrderRho12} \\
\varrho _{1,3}\left( t\right)  &=&-\Xi _{a}^{\left( 1\right) }\varrho
_{M_{ab}}\left( t_{0}\right) -\Xi _{b}^{\left( 1\right) }\varrho
_{N_{ab}}\left( t_{0}\right) +\Xi _{f}^{\left( 1\right) }\varrho
_{M_{fg}}\left( t_{0}\right) +\Xi _{g}^{\left( 1\right) }\varrho
_{N_{fg}}\left( t_{0}\right) \text{,} \label{SpinSystemAXFirstOrderRho13} \\
\varrho _{2,4}\left( t\right)  &=&+\Xi _{a}^{\left( 1\right) }\varrho
_{M_{ab}}\left( t_{0}\right) +\Xi _{b}^{\left( 1\right) }\varrho
_{N_{ab}}\left( t_{0}\right) +\Xi _{f}^{\left( 1\right) }\varrho
_{M_{fg}}\left( t_{0}\right) +\Xi _{g}^{\left( 1\right) }\varrho
_{N_{fg}}\left( t_{0}\right) \text{,} \label{SpinSystemAXFirstOrderRho24} \\
\varrho _{3,4}\left( t\right)  &=&+\Xi _{a}^{\left( 1\right) }\varrho
_{K_{ab}}\left( t_{0}\right) +\Xi _{b}^{\left( 1\right) }\varrho
_{L_{ab}}\left( t_{0}\right) +\Xi _{f}^{\left( 1\right) }\varrho
_{K_{fg}}\left( t_{0}\right) +\Xi _{g}^{\left( 1\right) }\varrho
_{L_{fg}}\left( t_{0}\right) \text{,} \label{SpinSystemAXFirstOrderRho34}
\end{eqnarray}
\end{subequations}
where $\Xi _{\xi }^{\left( 1\right) }$ means the exponential function
depending on the relaxation rates of first order as denoted by $\exp \left[
-R_{\xi }^{\left( 1\right) }\left( t-t_{0}\right) \right] $ with the
subscript $\xi =a,b,f,g$. The relaxation rate constants are denoted by  $ R _{\xi }^{\left( 1 \right) }=\mathcal{C} \ \iota_{\xi }^{\left( 1 \right) }  $, where $\iota_{\xi }^{\left( 1 \right) }$ represents any eigenvalue of the superoperator $\boldsymbol{\mathcal{J}}^{\left( 1\right)}$, and those can be resumed in Tab. \ref{tab:RelaxationRatesAX}. Furthermore, the coefficients multiplying each exponential function depend on an appropriate linear combination of the density matrix elements of the initial quantum state, and they are denoted as follows
\begin{subequations}
\label{SpinSystemAXFirstOrderCoefficients}
\begin{eqnarray}
\varrho _{K_{ab}}\left( t_{0}\right)  &=&+\frac{q_{1}p_{2}\left( \varrho
_{1,2}\left( t_{0}\right) -\varrho _{3,4}\left( t_{0}\right) \right) }{%
2p_{1}q_{2}-2p_{2}q_{1}}-\frac{q_{1}q_{2}\left( \varrho _{1,3}\left(
t_{0}\right) -\varrho _{2,4}\left( t_{0}\right) \right) }{%
2p_{1}q_{2}-2p_{2}q_{1}}\text{,} \quad \label{SpinSystemAXFirstOrderCoefficientsKab} \\
\varrho _{L_{ab}}\left( t_{0}\right)  &=&-\frac{p_{1}q_{2}\left( \varrho
_{1,2}\left( t_{0}\right) -\varrho _{3,4}\left( t_{0}\right) \right) }{%
2p_{1}q_{2}-2p_{2}q_{1}}+\frac{q_{1}q_{2}\left( \varrho _{1,3}\left(
t_{0}\right) -\varrho _{2,4}\left( t_{0}\right) \right) }{%
2p_{1}q_{2}-2p_{2}q_{1}}\text{,} \label{SpinSystemAXFirstOrderCoefficientsLab} \\
\varrho _{M_{ab}}\left( t_{0}\right)  &=&+\frac{p_{1}p_{2}\left( \varrho
_{1,2}\left( t_{0}\right) -\varrho _{3,4}\left( t_{0}\right) \right) }{%
2p_{1}q_{2}-2p_{2}q_{1}}-\frac{p_{1}q_{2}\left( \varrho _{1,3}\left(
t_{0}\right) -\varrho _{2,4}\left( t_{0}\right) \right) }{%
2p_{1}q_{2}-2p_{2}q_{1}}\text{,} \label{SpinSystemAXFirstOrderCoefficientsMab} \\
\varrho _{N_{ab}}\left( t_{0}\right)  &=&-\frac{p_{1}p_{2}\left( \varrho
_{1,2}\left( t_{0}\right) -\varrho _{3,4}\left( t_{0}\right) \right) }{%
2p_{1}q_{2}-2p_{2}q_{1}}+\frac{q_{1}p_{2}\left( \varrho _{1,3}\left(
t_{0}\right) -\varrho _{2,4}\left( t_{0}\right) \right) }{%
2p_{1}q_{2}-2p_{2}q_{1}}\text{,} \label{SpinSystemAXFirstOrderCoefficientsNab} \\
\varrho _{K_{fg}}\left( t_{0}\right)  &=&-\frac{q_{3}p_{4}\left( \varrho
_{1,2}\left( t_{0}\right) +\varrho _{3,4}\left( t_{0}\right) \right) }{%
2p_{3}q_{4}-2p_{4}q_{3}}+\frac{q_{3}q_{4}\left( \varrho _{1,3}\left(
t_{0}\right) +\varrho _{2,4}\left( t_{0}\right) \right) }{%
2p_{3}q_{4}-2p_{4}q_{3}}\text{,} \label{SpinSystemAXFirstOrderCoefficientsKfg} \\
\varrho _{L_{fg}}\left( t_{0}\right)  &=&+\frac{p_{3}q_{4}\left( \varrho
_{1,2}\left( t_{0}\right) +\varrho _{3,4}\left( t_{0}\right) \right) }{%
2p_{3}q_{4}-2p_{4}q_{3}}-\frac{q_{3}q_{4}\left( \varrho _{1,3}\left(
t_{0}\right) +\varrho _{2,4}\left( t_{0}\right) \right) }{%
2p_{3}q_{4}-2p_{4}q_{3}}\text{,} \label{SpinSystemAXFirstOrderCoefficientsLfg} \\
\varrho _{M_{fg}}\left( t_{0}\right)  &=&-\frac{p_{3}p_{4}\left( \varrho
_{1,2}\left( t_{0}\right) +\varrho _{3,4}\left( t_{0}\right) \right) }{%
2p_{3}q_{4}-2p_{4}q_{3}}+\frac{p_{3}q_{4}\left( \varrho _{1,3}\left(
t_{0}\right) +\varrho _{2,4}\left( t_{0}\right) \right) }{%
2p_{3}q_{4}-2p_{4}q_{3}}\text{,} \label{SpinSystemAXFirstOrderCoefficientsMfg} \\
\varrho _{N_{fg}}\left( t_{0}\right)  &=&+\frac{p_{3}p_{4}\left( \varrho
_{1,2}\left( t_{0}\right) +\varrho _{3,4}\left( t_{0}\right) \right) }{%
2p_{3}q_{4}-2p_{4}q_{3}}-\frac{q_{3}p_{4}\left( \varrho _{1,3}\left(
t_{0}\right) +\varrho _{2,4}\left( t_{0}\right) \right) }{%
2p_{3}q_{4}-2p_{4}q_{3}}\text{,} \label{SpinSystemAXFirstOrderCoefficientsNfg}
\end{eqnarray}
\end{subequations}
where the auxiliary parameters $p_{k}$ and $q_{k}$ are resumed by
\begin{eqnarray*}
p_{1} &=&+9\left( J_{S}-J_{I}\right) ^{2}-3\left( J_{S}-J_{I}\right) 4\left(
J_{-}+J_{0}\right) +16\left( J_{-}+J_{0}\right) ^{2} \\
&&-12\left( 3\left( J_{S}-J_{I}\right) -4\left( J_{0}+J_{-}\right) \right)
\Lambda \text{,} \\
q_{1} &=&16\left( J_{0}+J_{-}\right) \left( \left( J_{0}+J_{-}\right)
+3\Lambda \right) 
\end{eqnarray*}%
\begin{eqnarray*}
p_{2} &=&+9\left( J_{S}-J_{I}\right) ^{2}-12\left( J_{S}-J_{I}\right) \left(
J_{-}+J_{0}\right) +16\left( J_{-}+J_{0}\right) ^{2} \\
&&+12\left( 3\left( J_{S}-J_{I}\right) -4\left( J_{0}+J_{-}\right) \right)
\Lambda \text{,} \\
q_{2} &=&16\left( J_{0}+J_{-}\right) \left( \left( J_{0}+J_{-}\right)
-3\Lambda \right) \text{,}
\end{eqnarray*}%
\begin{eqnarray*}
p_{4} &=&+36J_{S}^{2}+27\left( J_{S}+J_{I}\right) ^{2}+12\left(
5J_{S}+3J_{I}\right) \left( J_{-}+J_{0}\right) +16\left( J_{-}+J_{0}\right)
^{2} \\
&&-12\left( 3\left( 3J_{S}+J_{I}\right) +4\left( J_{-}+J_{0}\right) \right)
\Psi \text{,} \\
q_{4} &=&\left( 6\left( J_{S}+J_{I}\right) +4\left( J_{-}+J_{0}\right)
\right) ^{2}-24\left( 3\left( J_{S}+J_{I}\right) +2\left( J_{-}+J_{0}\right)
\right) \Psi \text{,}
\end{eqnarray*}%
\begin{eqnarray*}
p_{3} &=&+36J_{S}^{2}+27\left( J_{S}+J_{I}\right) ^{2}+12\left(
5J_{S}+3J_{I}\right) \left( J_{-}+J_{0}\right) +16\left( J_{-}+J_{0}\right)
^{2} \\
&&+12\left( 3\left( 3J_{S}+J_{I}\right) +4\left( J_{-}+J_{0}\right) \right)
\Psi \text{,} \\
q_{3} &=&+\left( 6\left( J_{S}+J_{I}\right) +4\left( J_{-}+J_{0}\right)
\right) ^{2}+24\left( 3\left( J_{S}+J_{I}\right) +2\left( J_{-}+J_{0}\right)
\right) \Psi \text{,}
\end{eqnarray*}
with the the radical expressions resumed by
\begin{eqnarray*}
\Lambda  &=&\sqrt{\left( \frac{J_{S}-J_{I}}{4}\right) ^{2}+\left( \frac{%
J_{-}+J_{0}}{3}\right) ^{2}}\text{,} \\
\Psi  &=&\sqrt{\left( \frac{J_{S}-J_{I}}{4}\right) ^{2}+\left( \frac{%
J_{S}+J_{I}}{2}+\frac{J_{-}+J_{0}}{3}\right) ^{2}}\text{.}
\end{eqnarray*}

Finally, the solution generated by the single differential equation corresponds to the second-order density matrix element with the relaxation rate constant denoted by  $ R _{a}^{\left( 2 \right) }=\mathcal{C} \ \iota_{a}^{\left( 2 \right) }  $, where $\iota_{a}^{\left( 2 \right) }$ means the eigenvalue of the superoperator $\boldsymbol{\mathcal{J}}^{\left( 2\right)}$, and it can be resumed in Tab.  \ref{tab:RelaxationRatesAX}. Therefore,  the solution is written
\begin{equation}
\varrho _{1,4}\left( t\right) =\exp \left[ -R _{a}^{\left( 2\right) }\left( t-t_{0}\right) \right] \varrho _{1,4}\left( t_{0}\right) \text{.} \label{SpinSystemAXSecondOrderRho14}
\end{equation}

A figure of merit of this theoretical description stays on generating a linear system of six differential equations related to the zero-order density matrix elements. The solution of the linear system introduces a set of relaxation rate constants that are compatible with the actual ones (see Eq. (5.45) of Ref. \cite{kruk2016Book}, Eq. (22) of Ref. \cite{hausser1968}, and others \cite{abragam1994Book,mcconnell1987Book})).

An analogous development was performed in the case of the four (the single) differential equations related to the first-order (second-order) density matrix elements. Those solutions are showed, but they are not explored deeply in this study as done with the zero-order solutions.

\section{Applications}
\label{sec:Applications}

With the solutions in hand, we will discuss two applications. The first one application is devoted to the computation of the $z$- and $x$-components  mean values of the spin angular momentum operators representing an ensemble of interacting pairs of unlike spins at room temperature. The mathematical definition of the mean value of any operator ${\mathbf{O}}$  used in this analysis is $\left\langle {\mathbf{O}}\right\rangle = \text{\texttt{Tr}}\left\{{\mathbf{\varrho }}{\mathbf{O}}\right\}$ where the matrix notation of the density operator  ${\mathbf{\varrho}}$ is defined by Eq. (\ref{MatricialRepresentationDensityOperator}). The other  application is  devoted to simulating the evolution of the elements of the density matrix of an initial pure state evolving to another pure state.

\subsection{Longitudinal magnetization}
\label{sec:LongitudinalMagnetization}

The computation of the $z$-components angular momentum operator mean values of each nuclear species,  considering the steady-state of thermal equilibrium bound, are resumed as follows
\begin{eqnarray}
\left\langle {\mathbf{I}}_{z}\right\rangle &=&\frac{\varrho _{1,1}\left(
t\right) +\varrho _{2,2}\left( t\right) -\varrho _{3,3}\left( t\right)
-\varrho _{4,4}\left( t\right) }{2} \text{,} \nonumber \\
 &=&\left\langle {\mathbf{I}}%
_{z}\right\rangle _{\text{eq}}+\exp \left[ - \mathcal{C} \, \iota _{f}^{\left( 0\right) }\left( t-t_{0}\right) \right] \left\langle {\mathbf{I}}_{z}\right\rangle
_{+}+\exp \left[ - \mathcal{C} \, \iota _{g}^{\left( 0\right) }\left( t-t_{0}\right) \right]
\left\langle {\mathbf{I}}_{z}\right\rangle _{-}\text{,} \qquad \label{MzNuclearSpecieI}  
\end{eqnarray}
\begin{eqnarray}
\left\langle {\mathbf{S}}_{z}\right\rangle &=&\frac{\varrho _{1,1}\left(
t\right) -\varrho _{2,2}\left( t\right) +\varrho _{3,3}\left( t\right)
-\varrho _{4,4}\left( t\right) }{2} \text{,} \nonumber \\
&=&\left\langle {\mathbf{S}}%
_{z}\right\rangle _{\text{eq}}-\exp \left[ - \mathcal{C} \, \iota _{f}^{\left( 0\right) }\left(
t-t_{0}\right) \right] \left\langle {\mathbf{S}}_{z}\right\rangle
_{+}-\exp \left[ - \mathcal{C} \, \iota _{g}^{\left( 0\right) }\left( t-t_{0}\right) \right]
\left\langle {\mathbf{S}}_{z}\right\rangle _{-}\text{,}  \quad \quad  \label{MzNuclearSpecieS}
\end{eqnarray}
where the coefficients are written explicitly as
\begin{equation}
\left\langle {\mathbf{I}}_{z}\right\rangle _{\text{eq}} = \frac{\varrho_{1,1}^{\text{eq}}+\varrho _{2,2}^{\text{eq}}-\varrho _{3,3}^{\text{eq}}-\varrho _{4,4}^{\text{eq}}}{2} \text{,}    \label{IzEquilibrio}
\end{equation}
\begin{eqnarray}
\left\langle {\mathbf{I}}_{z}\right\rangle _{+} &=&\frac{\left(A+B-C-D\right) \left(A-B\right)}{4\left(AD -BC \right) } \left(-\delta \rho _{1,1}\left( t_{0}\right) +\delta \rho _{4,4}\left(t_{0}\right) \right) \nonumber \\
 & & -\frac{\left(A+B-C-D\right) \left( C-D\right) }{4\left( AD-BC\right) } \left(\delta \rho _{2,2}\left(t_{0}\right) -\delta \rho _{3,3}\left( t_{0}\right) \right)\text{,} \qquad \label{IzPositive}
\end{eqnarray}
\begin{eqnarray}
\left\langle {\mathbf{I}}_{z}\right\rangle _{-} &=&\frac{\left(A-B-C+D\right) \left( A+B\right) }{4\left( AD-BC\right) }\left(\delta \rho _{1,1}\left( t_{0}\right) -\delta \rho _{4,4}\left( t_{0}\right)\right) \nonumber \\
 & &  +\frac{\left(A-B-C+D\right)\left( C+D\right)  }{4\left( AD-BC\right) }\left(\delta \rho _{2,2}\left( t_{0}\right) -\delta \rho _{3,3}\left( t_{0}\right) \right)\text{,} \label{IzNegative}
\end{eqnarray}
\begin{equation}
\left\langle {\mathbf{S}}_{z}\right\rangle _{\text{eq}} = \frac{\varrho_{1,1}^{\text{eq}}-\varrho _{2,2}^{\text{eq}}+\varrho _{3,3}^{\text{eq} }-\varrho _{4,4}^{\text{eq}}}{2} \text{,}    \label{SzEquilibrio}
\end{equation}
\begin{eqnarray}
\left\langle {\mathbf{S}}_{z}\right\rangle _{+} &=&\frac{\left( A+B+C+D\right) \left( A-B\right) }{4\left( AD-BC\right)}\left(-\delta \rho _{1,1}\left( t_{0}\right) +\delta \rho _{4,4}\left( t_{0}\right)\right) \nonumber \\
 & & -\frac{\left( A+B+C+D\right) \left( C-D\right)  }{4\left( AD-BC\right) }\left( \delta \rho _{2,2}\left( t_{0}\right) -\delta \rho _{3,3}\left( t_{0}\right) \right) \text{,}  \label{SzPositive}
\end{eqnarray}
\begin{eqnarray}
\left\langle {\mathbf{S}}_{z}\right\rangle _{-} &=&\frac{\left( A-B+C-D\right) \left( A+B\right) }{4\left( AD-BC\right)}\left(\delta \rho _{1,1}\left( t_{0}\right) -\delta \rho _{4,4}\left( t_{0}\right) \right) \nonumber \\
 & &+ \frac{\left( A-B+C-D\right)  \left( C+D\right) }{4\left( AD-BC\right) }   \left( \delta \rho _{2,2}\left( t_{0}\right) -\delta \rho _{3,3}\left( t_{0}\right) \right)   \text{.} \label{SzNegative}
\end{eqnarray}
The $z$-components of magnetization are characterized by two exponential functions, as expected from the previous standard theoretical procedures \cite{abragam1994Book,mcconnell1987Book}. 

The exact solution performed in this analysis generates different superpositions of spectral density function for the relaxation rate constants on the $z$-components of magnetization. These new mathematical expressions  do not change the longitudinal relaxation times of previously measured data but  will give new insights into thermal fluctuations and correlation times. In addition, the coefficients have shown a dependence on the spectral density functions demonstrated analytically.

The longitudinal magnetization of Eq. (\ref{MzNuclearSpecieI}) and (\ref{MzNuclearSpecieS}) were analyzed following the Solomon approach at the high-temperature regime. In this case, the quantum steady-state can be denoted by the density matrix in terms of the partition function $\mathcal{Z}$, the environment thermal energy, which is proportional to the room temperature $T$. Performing these assumptions, the density operator and their   non-null elements are written as follows
\begin{equation}
{\mathbf{\varrho }}^{\text{eq}}\approx \frac{1}{\mathcal{Z}}{\mathbf{%
1}}-\frac{1}{\mathcal{Z}} \frac{{\mathcal{H}}}{k_{B}T}\text{,}\qquad \Rightarrow
\qquad \left\{ 
\begin{array}{c}
\varrho _{1,1}^{\text{eq}}\approx \frac{1}{\mathcal{Z}}\left( 1+\frac{\epsilon ^{I}}{2}+\frac{ \epsilon ^{S}}{2}\right) \text{,} \\ 
\varrho _{2,2}^{\text{eq}}\approx \frac{1}{\mathcal{Z}}\left( 1+\frac{
\epsilon ^{I}}{2 }-\frac{ \epsilon ^{S}}{2 }\right) \text{,} \\ 
\varrho _{3,3}^{\text{eq}}\approx \frac{1}{\mathcal{Z}}\left( 1-\frac{
\epsilon ^{I}}{2 }+\frac{\epsilon ^{S}}{2 }\right) \text{,} \\ 
\varrho _{4,4}^{\text{eq}}\approx \frac{1}{\mathcal{Z}}\left( 1-\frac{
\epsilon ^{I}}{2 }-\frac{\epsilon ^{S}}{2 }\right) \text{,}%
\end{array}%
\right.  \label{QuantumStateEquilibrium}
\end{equation}
where $\epsilon^{S,I}$ mean the polarization parameter \cite{consuelo-leal2023JMR,dieguez2022,consuelo-leal2019,auccaise2011B,cius2022} or Boltzmann factor \cite{levitt2001Book,kowalewski2018Book} for both nuclear species, which were denoted by $\epsilon^{S,I} = \frac{  \hbar \omega ^{S,I}}{k_{B}T}$. Next, evaluating the mean value of the z-magnetization at the steady-state
\begin{eqnarray}
\left\langle {\mathbf{I}}_{z}\right\rangle _{\text{eq}} &=&\frac{\varrho 
_{1,1}^{\text{eq}}+\varrho _{2,2}^{\text{eq}}-\varrho _{3,3}^{\text{eq}%
}-\varrho _{4,4}^{\text{eq}}}{2}=\frac{ \epsilon ^{I}}{\mathcal{Z} } \label{MagnetizationZrhoEquilibrioEspecieI} \text{,} \\
\left\langle {\mathbf{S}}_{z}\right\rangle _{\text{eq}} &=&\frac{\varrho
_{1,1}^{\text{eq}}-\varrho _{2,2}^{\text{eq}}+\varrho _{3,3}^{\text{eq}%
}-\varrho _{4,4}^{\text{eq}}}{2}=\frac{\epsilon ^{S}}{\mathcal{Z} } \label{MagnetizationZrhoEquilibrioEspecieS}\text{,}
\end{eqnarray}
such that these results were substituted at the z-magnetization of both nuclear species 
\begin{eqnarray}
\left\langle {\mathbf{I}}_{z}\right\rangle  &=&\frac{ \epsilon ^{I}}{\mathcal{Z} }+\exp \left[ - \mathcal{C} \, \iota _{f}^{\left( 0\right) }\left(
t-t_{0}\right) \right] \left\langle {\mathbf{I}}_{z}\right\rangle
_{+}+\exp \left[ - \mathcal{C} \, \iota _{g}^{\left( 0\right) }\left( t-t_{0}\right) \right]
\left\langle {\mathbf{I}}_{z}\right\rangle _{-}\text{,} \label{MzNuclearSpecieI-Aplicar}   \\
\left\langle {\mathbf{S}}_{z}\right\rangle &=&\frac{ \epsilon ^{S}}{\mathcal{Z} }-\exp \left[ - \mathcal{C} \, \iota _{f}^{\left( 0\right) }\left(
t-t_{0}\right) \right] \left\langle {\mathbf{S}}_{z}\right\rangle
_{+}-\exp \left[ - \mathcal{C} \, \iota _{g}^{\left( 0\right) }\left( t-t_{0}\right) \right]
\left\langle {\mathbf{S}}_{z}\right\rangle _{-}\text{,}  \quad \quad  \label{MzNuclearSpecieS-Aplicar}
\end{eqnarray}
and these  magnetization expressions will be analyzed using three assumptions established by the initial quantum state, as discussed in the following subsections.

\subsection{Saturation magnetization of the $S$ nuclei}
\label{sec:SaturationMagnetizationSNuclearSpecie}

In this first setup, the saturation population of the $S$ nuclei must be performed \cite{abragam1994Book}. To achieve this task, the initial quantum state can be implemented by amplitude equalization between two density matrix elements \cite{cius2022,bull1992,viola2000}. After the execution of the saturation process, the initial density matrix of the spin system is denoted by
\begin{equation}
{\mathbf{\varrho }}\left( t_{0}\right) \approx \frac{1}{\mathcal{Z}}{%
\mathbf{1}}+\frac{ \epsilon  ^{I}}{2\mathcal{Z} }\left[ 
\begin{array}{cccc}
1 & 0 & 0 & 0 \\ 
0 & 1 & 0 & 0 \\ 
0 & 0 & -1 & 0 \\ 
0 & 0 & 0 & -1%
\end{array}%
\right] \  \Longrightarrow \  \left\{ 
\begin{array}{c}
\varrho _{1,1}\left( t_{0}\right) \approx \frac{1}{\mathcal{Z}}\left( 1+%
\frac{ \epsilon  ^{I}}{2 }\right) \text{,} \\ 
\varrho _{2,2}\left( t_{0}\right) \approx \frac{1}{\mathcal{Z}}\left( 1+%
\frac{ \epsilon  ^{I}}{2 }\right) \text{,} \\ 
\varrho _{3,3}\left( t_{0}\right) \approx \frac{1}{\mathcal{Z}}\left( 1-%
\frac{ \epsilon  ^{I}}{2 }\right) \text{,} \\ 
\varrho _{4,4}\left( t_{0}\right) \approx \frac{1}{\mathcal{Z}}\left( 1-%
\frac{ \epsilon  ^{I}}{2 }\right) \text{.}%
\end{array}%
\right.   \label{QuantumStateSaturation}
\end{equation}
Next, the explicit representation of the density matrix elements of the thermal equilibrium state (\ref{QuantumStateEquilibrium}) and the initial quantum state (\ref{QuantumStateSaturation}) allow  the computation of the coefficients defined in Eq. (\ref{IzPositive}), (\ref{IzNegative}), (\ref{SzPositive}), and (\ref{SzNegative}). After performing some standard algebraic procedures,  they are denoted by 
\begin{eqnarray*}
\left\langle {\mathbf{I}}_{z}\right\rangle _{+} &=&+\left( \frac{\epsilon ^{S}}{\mathcal{Z} }\right) \left( \frac{\left( A-C\right)
^{2}-\left( B-D\right) ^{2}}{4\left( AD-BC\right) }\right) \text{,} \\
\left\langle {\mathbf{I}}_{z}\right\rangle _{-} &=&-\left( \frac{\epsilon ^{S}}{\mathcal{Z} }\right) \left( \frac{\left( A-C\right)
^{2}-\left( B-D\right) ^{2}}{4\left( AD-BC\right) }\right) \text{,} \\%
\left\langle {\mathbf{S}}_{z}\right\rangle _{+} &=&+\left( \frac{\epsilon ^{S}}{\mathcal{Z} }\right) \left( \frac{A^{2}-B^{2}-C^{2}+D^{2}}{%
4\left( AD-BC\right) }+\frac{1}{2}\right) \text{,} \\
\left\langle {\mathbf{S}}_{z}\right\rangle _{-} &=&-\left( \frac{\epsilon ^{S}}{\mathcal{Z} }\right) \left( \frac{A^{2}-B^{2}-C^{2}+D^{2}}{%
4\left( AD-BC\right) }-\frac{1}{2}\right) \text{.}
\end{eqnarray*}

Besides the computation of those four coefficients, they must be substituted on the z-magnetization definition introduced by Eq. (\ref {MzNuclearSpecieI-Aplicar}) and (\ref{MzNuclearSpecieS-Aplicar}) of both species 
\begin{eqnarray*}
\left\langle {\mathbf{I}}_{z}\right\rangle  &=&\left( \frac{\epsilon ^{I}}{\mathcal{Z} }\right) -\left( \frac{\epsilon ^{S}}{\mathcal{Z} }\right) \left( h_{1}\right) \left( \exp \left[ -\mathcal{C} \,\iota _{g}^{\left( 0\right) }\left( t-t_{0}\right) \right] -\exp \left[ -%
\mathcal{C}\,\iota _{f}^{\left( 0\right) }\left( t-t_{0}\right) \right]
\right) \text{,} \\
\left\langle {\mathbf{S}}_{z}\right\rangle  &=&\left( \frac{\epsilon ^{S}}{\mathcal{Z} }\right) +\left( \frac{\epsilon ^{S}}{\mathcal{Z} }\right) \left( h_{2}\right) \left( \exp \left[ -\mathcal{C}%
\,\iota _{g}^{\left( 0\right) }\left( t-t_{0}\right) \right] -\exp \left[ -%
\mathcal{C}\,\iota _{f}^{\left( 0\right) }\left( t-t_{0}\right) \right]
\right)  \\
&&-\left( \frac{\epsilon ^{S}}{\mathcal{Z} }\right) \left( \frac{1}{%
2}\right) \left( \exp \left[ -\mathcal{C}\,\iota _{g}^{\left( 0\right)
}\left( t-t_{0}\right) \right] +\exp \left[ -\mathcal{C}\,\iota _{f}^{\left(
0\right) }\left( t-t_{0}\right) \right] \right) \text{,}
\end{eqnarray*}
\begin{SCfigure}
\includegraphics[width=3.0in]{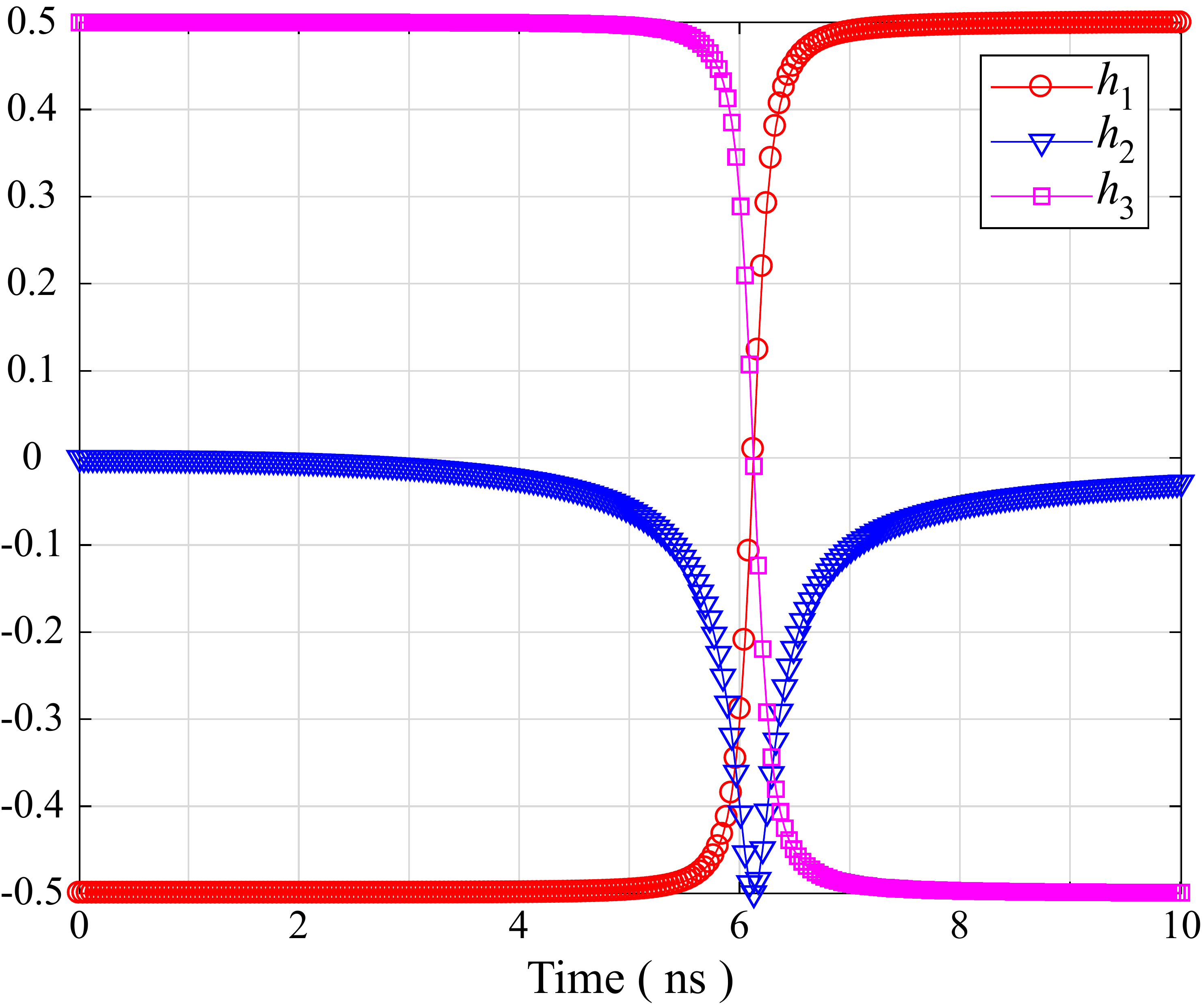} 
\caption{(Color online) Illustration of the three coefficients as denoted in Eq.(\ref{eq:CoefficientH1}-\ref{eq:CoefficientH3}), represented by the circle, triangle and square symbols. The computation of those data was predicted for an anhydrous hydrofluoric acid molecule submitted under the magnetic field strength of $0.705$ T and the ``isotropic model'' for the spectral density functions in terms of the  correlation time  $\tau_{c}$.} \label{fig:AuxiliaryFunctionsForMagnetization}
\end{SCfigure}
where the coefficients $h_{1}$ and $h_{2}$ must be introduced as a shorthand notation for some other ones that depend on $A$, $B$, $C$, and $D$ functions and explicitly  denoted by 
\begin{eqnarray}
h_{1} &=&\frac{\left( A-C\right) ^{2}-\left( B-D\right) ^{2}}{4\left(
AD-BC\right) }\text{,}  \label{eq:CoefficientH1} \\
h_{2} &=&\frac{A^{2}-B^{2}-C^{2}+D^{2}}{4\left( AD-BC\right) }\text{,}
\label{eq:CoefficientH2} \\
h_{3} &=&\frac{\left( A+C\right) ^{2}-\left( B+D\right) ^{2}}{4\left(
AD-BC\right) }\text{.}  \label{eq:CoefficientH3}
\end{eqnarray}%
Each of those three coefficients can be written using Eq. (\ref{eq:NormaTransParam}), and they save their dependence on the spectral density function. For example, let us consider the anhydrous hydrofluoric acid at a static magnetic field strength of $0.705$ T, as detailed in Ref. \cite{bloembergen1948}, also consider the ``isotropic model'' for the spectral density functions defined by \cite{abragam1994Book,bull1972}
\begin{equation}
J\left( \omega \right) = \frac{2\tau }{1+\left( \omega \tau  \right) ^{2}} \text{,} \label{ModelIsotropic}
\end{equation}
where $ \tau$ represents the correlation time that characterizes molecular motions (as was detailed in the description of Eq.  (\ref{EquationMotionByRedfieldTheory})) and $\omega$ represents the characteristic angular frequency of an eigen-energy of the system. For Hydrogen and Fluorine nuclei, the values of the angular frequency were calculated and denoted by $\omega_{S}=188.603\times 10^{6}$Hz and  $\omega_{I}=177.394\times 10^{6}$Hz, respectively. Assuming the correlation time as an independent variable, those coefficients have their smooth variation graphically displayed in Fig. \ref{fig:AuxiliaryFunctionsForMagnetization}. 

These pictorial representations support the predicted value $1/2$ found by I. Solomon because following the solution solving the Redfield equation, the coefficient is $-h_{1}=1/2$. An analogous argument is devoted using the ``Model-free'' for the spectral density functions as introduced by Lipari-Szabo \cite{lipari1982A}, as denoted by
\begin{equation}
J\left( \omega \right) =\frac{2}{5}\left( \frac{S^{2}\tau _{M}}{1+\left(
\omega \tau _{M}\right) ^{2}}+\frac{\left( 1-S^{2}\right) \tau }{1+\left(
\omega \tau \right) ^{2}}\right) \text{,\qquad }\frac{1}{\tau }=\frac{1}{%
\tau _{M}}+\frac{1}{\tau _{e}}\text{,}
\end{equation}
where $ \tau _{M}$ represents the correlation time that characterizes macro-molecular motions,  $ \tau _{e}$ represents the correlation time of effective motions, and $S$ represents a generalized order parameter (measures the degree of spatial restriction of the motion). Even with these more sophisticated spectral density functions, the coefficient reproduces the same value  $-h_{1}=1/2$.

Considering the initial time $t_{0}=0$ and the definition of the magnetization at the steady-state as denoted by Eq. (\ref{MagnetizationZrhoEquilibrioEspecieI}) and (\ref{MagnetizationZrhoEquilibrioEspecieS}), the angular momentum operators' mean values are resumed by 
\begin{eqnarray}
\frac{\left\langle {\mathbf{I}}_{z}\right\rangle -\left\langle {\mathbf{I}}_{z}\right\rangle _{\text{eq}}}{\left\langle {\mathbf{S}} _{z}\right\rangle _{\text{eq}}} \ =\ \Upsilon _{\text{Sat}}^{I}\left( t\right)
 &=& -h_{1}\left( \exp \left[ -\mathcal{C}\,\iota _{g}^{\left( 0\right) }t \right] -\exp \left[ -\mathcal{C}\,\iota _{f}^{\left( 0\right) }t\right] \right) \text{,}  \label{MzNuclearSpecieI-1roQuanState} \\
\frac{\left\langle {\mathbf{S}}_{z}\right\rangle -\left\langle { \mathbf{S}}_{z}\right\rangle _{\text{eq}}}{\left\langle {\mathbf{S}} _{z}\right\rangle _{\text{eq}}} \ =\ \Upsilon _{\text{Sat}}^{S}\left( t\right)
 &=&  h_{2}\left( \exp \left[ -\mathcal{C}\,\iota _{g}^{\left( 0\right) }t \right] -\exp \left[ -\mathcal{C}\,\iota _{f}^{\left( 0\right) }t\right] \right) \nonumber \\ 
&&-\frac{1}{2}\left( \exp \left[ -\mathcal{C}\,\iota _{g}^{\left( 0\right) }t\right] +\exp \left[ -\mathcal{C}\,\iota _{f}^{\left( 0\right) }t \right] \right) \text{,}  \label{MzNuclearSpecieS-1roQuanState}
\end{eqnarray}
where we use $\Upsilon _{\text{Sat}}^{I}\left( t\right) $ and $\Upsilon _{\text{Sat}}^{S}\left( t\right) $ as a concise notation of Eq. (44) in Ref.  \cite{solomon1955} introduced by I. Solomon. These magnetization functions will help us to highlight the spin dynamics and compare them at the same scale frame.

\subsection{Inversion magnetization of the $S$ nuclei}
\label{sec:InversionMagnetizationSNuclearSpecie}

In this second setup, the inversion population of the $S$ nuclei must be performed. The initial quantum state is implemented by transforming the steady-state using a unitary rotation named $\pi$-pulse \cite{abragam1994Book}. Ensuring the implementation of the inversion population, the initial density matrix of the spin system is denoted by 
\begin{equation}
{\mathbf{\varrho }}\left( t_{0}\right) \approx \frac{1}{\mathcal{Z}}{%
\mathbf{1}}+\frac{ \left( \epsilon ^{S}+\epsilon ^{I}\right) }{2%
\mathcal{Z} } {\mathbf{\Delta\varrho }} \text{,\quad }\Longrightarrow \quad \left\{ 
\begin{array}{c}
\varrho _{1,1}\left( t_{0}\right) \approx \frac{1}{\mathcal{Z}}\left( 1+%
\frac{ \left(\epsilon ^{I}-\epsilon ^{S}\right) }{2 }\right) \text{,}
\\ 
\varrho _{2,2}\left( t_{0}\right) \approx \frac{1}{\mathcal{Z}}\left( 1+%
\frac{ \left(\epsilon ^{I}+\epsilon ^{S}\right) }{2 }\right) \text{,}
\\ 
\varrho _{3,3}\left( t_{0}\right) \approx \frac{1}{\mathcal{Z}}\left( 1-%
\frac{ \left(\epsilon ^{I}+\epsilon ^{S}\right) }{2 }\right) \text{,}
\\ 
\varrho _{4,4}\left( t_{0}\right) \approx \frac{1}{\mathcal{Z}}\left( 1-%
\frac{ \left(\epsilon ^{I}-\epsilon ^{S}\right) }{2 }\right) \text{,}%
\end{array}%
\right.   \label{QuantumStateInversion}
\end{equation}
where the deviation matrix ${\mathbf{\Delta\varrho }}$ is denoted by
\begin{equation*}
{\mathbf{\Delta\varrho }}=\left[ 
\begin{array}{cccc}
+\frac{\omega ^{I}-\omega ^{S}}{\omega ^{I}+\omega ^{S}} & 0 & 0 & 0 \\ 
0 & 1 & 0 & 0 \\ 
0 & 0 & -1 & 0 \\ 
0 & 0 & 0 & -\frac{\omega ^{I}-\omega ^{S}}{\omega ^{I}+\omega ^{S}}%
\end{array}%
\right] \text{.}
\end{equation*}

Next, the explicit representation of the density matrix elements of the thermal equilibrium state (\ref{QuantumStateEquilibrium}) and the initial quantum state (\ref{QuantumStateInversion}) allow the computation of the coefficients defined in Eq. (\ref{IzPositive}), (\ref{IzNegative}), (\ref{SzPositive}), and (\ref{SzNegative}). After performing some standard algebraic procedures,  the mathematical expression of those coefficients can be denoted by 
\begin{eqnarray*}
\left\langle {\mathbf{I}}_{z}\right\rangle _{+} &=&-\left( \frac{2\epsilon ^{S}}{\mathcal{Z} }\right) \left( \frac{\left( B-D\right)
^{2}-\left( A-C\right) ^{2}}{4\left( AD-BC\right) }\right) \text{,} \\
\left\langle {\mathbf{I}}_{z}\right\rangle _{-} &=&+\left( \frac{2\epsilon ^{S}}{\mathcal{Z} }\right) \left( \frac{\left( B-D\right)
^{2}-\left( A-C\right) ^{2}}{4\left( AD-BC\right) }\right) \text{,} \\
\left\langle {\mathbf{S}}_{z}\right\rangle _{+} &=&+\left( \frac{%
2\epsilon ^{S}}{\mathcal{Z} }\right) \left( \frac{\left( A+D\right)
^{2}-\left( B+C\right) ^{2}}{4\left( AD-BC\right) }\right) \text{,} \\
\left\langle {\mathbf{S}}_{z}\right\rangle _{-} &=&+\left( \frac{%
2\epsilon ^{S}}{\mathcal{Z} }\right) \left( \frac{-\left(
A-D\right) ^{2}+\left( B-C\right) ^{2}}{4\left( AD-BC\right) }\right) \text{.}
\end{eqnarray*}
Besides the computation of those four coefficients, they must be substituted on the z-magnetization definition introduced by Eq. (\ref {MzNuclearSpecieI-Aplicar}) and (\ref{MzNuclearSpecieS-Aplicar}) of both species
\begin{eqnarray*}
\left\langle {\mathbf{I}}_{z}\right\rangle  &=&\left( \frac{\epsilon ^{I}}{\mathcal{Z} }\right) -\left( \frac{\epsilon ^{S}}{\mathcal{Z} }\right) \left( 2h_{1}\right) \left( \exp \left[ -\mathcal{C}\,\iota _{g}^{\left( 0\right) }\left( t-t_{0}\right) \right] -\exp \left[ -%
\mathcal{C}\,\iota _{f}^{\left( 0\right) }\left( t-t_{0}\right) \right]
\right) \text{,} \\
\left\langle {\mathbf{S}}_{z}\right\rangle  &=&\left( \frac{\epsilon ^{S}}{\mathcal{Z} }\right) +\left( \frac{\epsilon ^{S}}{\mathcal{Z} }\right) \left( 2h_{2}\right) \left( \exp \left[ -\mathcal{C}\,\iota _{g}^{\left( 0\right) }\left( t-t_{0}\right) \right] -\exp \left[ -%
\mathcal{C}\,\iota _{f}^{\left( 0\right) }\left( t-t_{0}\right) \right]
\right)  \\
&&-\left( \frac{\epsilon ^{S}}{\mathcal{Z} }\right) \left( \exp %
\left[ -\mathcal{C}\,\iota _{g}^{\left( 0\right) }\left( t-t_{0}\right) %
\right] +\exp \left[ -\mathcal{C}\,\iota _{f}^{\left( 0\right) }\left(
t-t_{0}\right) \right] \right) \text{,}
\end{eqnarray*}
where the coefficients $h_{1}$ and $h_{2}$ were defined in Eq. (\ref{eq:CoefficientH1}) and (\ref{eq:CoefficientH2}), respectively. Considering the initial time $t_{0}=0$ and the definition of the magnetization at the steady-state as denoted by Eq. (\ref{MagnetizationZrhoEquilibrioEspecieI}) and (\ref{MagnetizationZrhoEquilibrioEspecieS}), then the mean values of the angular momentum operators can be resumed by 
\begin{eqnarray}
\frac{\left\langle {\mathbf{I}}_{z}\right\rangle -\left\langle {%
\mathbf{I}}_{z}\right\rangle _{\text{eq}}}{\left\langle {\mathbf{S}}%
_{z}\right\rangle _{\text{eq}}} \ =\  \Upsilon _{\text{Inv}}^{I}\left( t\right)
&=&-2h_{1}\left( \exp \left[ -\mathcal{C}\,\iota _{g}^{\left( 0\right) }t%
\right] -\exp \left[ -\mathcal{C}\,\iota _{f}^{\left( 0\right) }t\right]
\right) \text{,}  \label{MzNuclearSpecieI-2doQuanState} \\
\frac{\left\langle {\mathbf{S}}_{z}\right\rangle -\left\langle {%
\mathbf{S}}_{z}\right\rangle _{\text{eq}}}{\left\langle {\mathbf{S}}%
_{z}\right\rangle _{\text{eq}}} \ =\ \Upsilon _{\text{Inv}}^{S}\left( t\right)
&=& +2h_{2}\left( \exp \left[ -\mathcal{C}\,\iota _{g}^{\left( 0\right) }t%
\right] -\exp \left[ -\mathcal{C}\,\iota _{f}^{\left( 0\right) }t\right]
\right) \nonumber \\
&& -\left( \exp \left[ -\mathcal{C}\,\iota _{g}^{\left( 0\right) }t%
\right] +\exp \left[ -\mathcal{C}\,\iota _{f}^{\left( 0\right) }t\right]
\right) \text{.} \qquad \label{MzNuclearSpecieS-2doQuanState}
\end{eqnarray}

\subsection{Inversion magnetization of both nuclei}
\label{sec:InversionMagnetizationBothNuclearSpecies}

In this third setup, the inversion population of the $I$ and $S$ nuclei must be performed. As in the previous case, the initial quantum state is implemented by transforming the steady-state using a unitary rotation, named $\pi$-pulse and carried out on both nuclear species \cite{abragam1994Book}. After establishing  the implementation of the inversion population, then the initial density matrix of the spin system is denoted by
\begin{equation}
{\mathbf{\varrho }}\left( t_{0}\right) \approx \frac{1}{\mathcal{Z}}{%
\mathbf{1}}+\frac{ \left(\epsilon ^{S}+\epsilon ^{I}\right) }{2%
\mathcal{Z} } {\mathbf{\Delta\varrho }} \quad \Longrightarrow \quad \left\{ 
\begin{array}{c}
\varrho _{1,1}\left( t_{0}\right) \approx \frac{1}{\mathcal{Z}}\left( 1-%
\frac{ \left(\epsilon ^{I}+\epsilon ^{S}\right) }{2 }\right) \text{,}
\\ 
\varrho _{2,2}\left( t_{0}\right) \approx \frac{1}{\mathcal{Z}}\left( 1-%
\frac{ \left(\epsilon ^{I}-\epsilon ^{S}\right) }{2 }\right) \text{,}
\\ 
\varrho _{3,3}\left( t_{0}\right) \approx \frac{1}{\mathcal{Z}}\left( 1+%
\frac{ \left(\epsilon ^{I}-\epsilon ^{S}\right) }{2 }\right) \text{,}
\\ 
\varrho _{4,4}\left( t_{0}\right) \approx \frac{1}{\mathcal{Z}}\left( 1+%
\frac{ \left(\epsilon ^{I}+\epsilon ^{S}\right) }{2 }\right) \text{,}%
\end{array}%
\right.   \label{QuantumStateInversionOnBoth}
\end{equation}
where the deviation matrix ${\mathbf{\Delta\varrho }}$ is denoted by 
\begin{equation*}
{\mathbf{\Delta\varrho }}=\left[ 
\begin{array}{cccc}
-1 & 0 & 0 & 0 \\ 
0 & -\frac{\omega ^{I}-\omega ^{S}}{\omega ^{S}+\omega ^{I}} & 0 & 0 \\ 
0 & 0 & +\frac{\omega ^{I}-\omega ^{S}}{\omega ^{S}+\omega ^{I}} & 0 \\ 
0 & 0 & 0 & +1%
\end{array}%
\right] \text{.}
\end{equation*}
Next, the explicit representation of the density matrix elements of the thermal equilibrium state (\ref{QuantumStateEquilibrium}) and the initial quantum state (\ref{QuantumStateInversionOnBoth}) allow the computation of the coefficients defined in Eq. (\ref{IzPositive}), (\ref{IzNegative}), (\ref{SzPositive}), and (\ref{SzNegative}). After performing some standard algebraic procedures and using Eq. (\ref{eq:CoefficientH1}-\ref{eq:CoefficientH3}),  the mathematical expression of those coefficients can be denoted by
\begin{eqnarray*}
\left\langle {\mathbf{I}}_{z}\right\rangle _{+} &=&+\left( \frac{2 
\epsilon ^{S}}{\mathcal{Z} }\right) \left( h_{1}\right) +\left( \frac{%
2 \epsilon ^{I}}{\mathcal{Z} }\right) \left( h_{2}-\frac{1}{2}\right) \text{,} \\
\left\langle {\mathbf{I}}_{z}\right\rangle _{-} &=&-\left( \frac{2 
\epsilon ^{S}}{\mathcal{Z} }\right) \left( h_{1}\right) -\left( \frac{%
2 \epsilon ^{I}}{\mathcal{Z} }\right) \left( h_{2}+\frac{1}{2}\right) \text{,} \\
\left\langle {\mathbf{S}}_{z}\right\rangle _{+} &=&+\left( \frac{2 
\epsilon ^{S}}{\mathcal{Z} }\right) \left( h_{2}+\frac{1}{2}\right) +\left( \frac{2 \epsilon 
^{I}}{\mathcal{Z} }\right) \left( h_{3}\right) \text{,} \\
\left\langle {\mathbf{S}}_{z}\right\rangle _{-} &=&-\left( \frac{2 
\epsilon ^{S}}{\mathcal{Z} }\right) \left( h_{2}-\frac{1}{2}\right) -\left( \frac{2 \epsilon 
^{I}}{\mathcal{Z} }\right) \left( h_{3}\right) \text{.}
\end{eqnarray*}
Besides the computation of those four coefficients, they must be substituted on the z-magnetization definition introduced by Eq. (\ref {MzNuclearSpecieI-Aplicar}) and (\ref{MzNuclearSpecieS-Aplicar}) of both species
\begin{eqnarray}
\left\langle {\mathbf{I}}_{z}\right\rangle  &=&\left( \frac{\epsilon ^{I}}{\mathcal{Z} }\right) -\left( 2h_{1}\left( \frac{\epsilon ^{S}}{\mathcal{Z} }\right) +2h_{2}\left( \frac{\epsilon ^{I}%
}{\mathcal{Z} }\right) \right) \left( \exp \left[ -\mathcal{C}\,\iota
_{g}^{\left( 0\right) }\left( t-t_{0}\right) \right]  \right)  
\notag \\
  & &  +\left( 2h_{1}\left( \frac{\epsilon ^{S}}{\mathcal{Z} }\right) +2h_{2}\left( \frac{\epsilon ^{I}}{\mathcal{Z} }\right) \right) \left(\exp \left[ -\mathcal{C%
}\,\iota _{f}^{\left( 0\right) }\left( t-t_{0}\right) \right] \right)  
\notag \\
&&-\left( \exp \left[ -\mathcal{C}\,\iota _{f}^{\left( 0\right) }\left(
t-t_{0}\right) \right] +\exp \left[ -\mathcal{C}\,\iota _{g}^{\left(
0\right) }\left( t-t_{0}\right) \right] \right) \left( \frac{\epsilon 
 ^{I}}{\mathcal{Z} }\right) \text{,} \\
\left\langle {\mathbf{S}}_{z}\right\rangle  &=&\left( \frac{\epsilon ^{S}}{\mathcal{Z} }\right) -\exp \left[ -\mathcal{C}\,\iota
_{f}^{\left( 0\right) }\left( t-t_{0}\right) \right] \left( \left[ 2h_{2}+1%
\right] \left( \frac{\epsilon ^{S}}{\mathcal{Z} }\right)
+2h_{3}\left( \frac{\epsilon ^{I}}{\mathcal{Z} }\right) \right) \notag \\
&&-\exp \left[ -\mathcal{C}\,\iota _{g}^{\left( 0\right) }\left(
t-t_{0}\right) \right] \left( -\left[ 2h_{2}-1\right] \left( \frac{ 
\epsilon ^{S}}{\mathcal{Z} }\right) -2h_{3}\left( \frac{\epsilon 
 ^{I}}{\mathcal{Z} }\right) \right) \text{,}
\end{eqnarray}%
where the coefficients $h_{1}$, $h_{2}$, and $h_{3}$ are defined in Eq. (\ref%
{eq:CoefficientH1}), (\ref{eq:CoefficientH2}), and (\ref{eq:CoefficientH3}),
respectively. Considering the initial time $t_{0}=0$ and the definition of
the magnetization at the steady state as denoted by Eq. (\ref%
{MagnetizationZrhoEquilibrioEspecieI}) and (\ref%
{MagnetizationZrhoEquilibrioEspecieS}), the mean values of the angular
momentum operators are resumed by 
\begin{eqnarray}
\left. \frac{\left\langle {\mathbf{I}}_{z}\right\rangle -\left\langle 
{\mathbf{I}}_{z}\right\rangle _{\text{eq}}}{\left\langle {\mathbf{S}}%
_{z}\right\rangle _{\text{eq}}}=\Upsilon _{\text{Inv-Both}}^{I}\left(
t\right) \right.  &=&-2\left( h_{1}+h_{2}\frac{\omega ^{I}}{\omega ^{S}}%
\right) \left( \exp \left[ -\mathcal{C}\,\iota _{g}^{\left( 0\right) }t%
\right] 
\right)   \notag \\
 & &+2\left( h_{1}+h_{2}\frac{\omega ^{I}}{\omega ^{S}}%
\right) \left( \exp \left[ -\mathcal{C}\,\iota _{f}^{\left( 0\right) }t\right]
\right)   \notag \\
&&-\frac{\omega ^{I}}{\omega ^{S}}\left( \exp \left[ -\mathcal{C}\,\iota
_{g}^{\left( 0\right) }t\right] +\exp \left[ -\mathcal{C}\,\iota
_{f}^{\left( 0\right) }t\right] \right) \text{,}
\label{MzNuclearSpecieI-3rdQuanState} \\
\left. \frac{\left\langle {\mathbf{S}}_{z}\right\rangle -\left\langle 
{\mathbf{S}}_{z}\right\rangle _{\text{eq}}}{\left\langle {\mathbf{S}}%
_{z}\right\rangle _{\text{eq}}}=\Upsilon _{\text{Inv-Both}}^{S}\left(
t\right) \right.  &=&-\exp \left[ -\mathcal{C}\,\iota _{f}^{\left( 0\right)
}t\right] \left( \left( 2h_{2}+1\right) +2h_{3}\frac{\omega ^{I}}{\omega ^{S}%
}\right)   \notag \\
&&+\exp \left[ -\mathcal{C}\,\iota _{g}^{\left( 0\right) }t\right] \left(
\left( 2h_{2}-1\right) +2h_{3}\frac{\omega ^{I}}{\omega ^{S}}\right) \text{.} \quad
\label{MzNuclearSpecieS-3rdQuanState}
\end{eqnarray}

These last three subsections demonstrated how the magnetization dynamics change under the initial quantum state choice. It is essential to highlight that the common physical characteristic of these theoretical results is that the magnetization evolution depends on the relaxation rate constants denoted by $ R_{f}^{\left( 0\right) } = \mathcal{C}\,\iota _{f}^{\left( 0\right) }$ and $R_{g}^{\left( 0\right) } = \mathcal{C}\,\iota _{g}^{\left( 0\right) }$, which are compatible with other previous predictions performed by applying the Redfield equations for operators mean values \cite{kruk2016Book}. It warrants the solutions of the density matrix elements preserve previous theoretical findings as measured observables of standard theoretical procedures and experimental techniques \cite{cavanagh2007Book,kruk2016Book,kowalewski2002,canet2011,jaroszkiewicz1985,koenig1990}.

\subsection{Transverse magnetization}
\label{sec:TransversalMagnetization}

The computation of the $x$-components angular momentum operator mean values of each nuclei species, and considering the initial quantum state at the time $t_{0}$, are resumed as follows 
\begin{eqnarray}
\left\langle {\mathbf{I}}_{x}\right\rangle &=&\frac{1}{2}\left( \varrho
_{1,3}+\varrho _{2,4}+\varrho _{3,1} +\varrho _{4,2} \right) =%
\frac{1}{2}\left( \left( \varrho _{1,3}+\varrho _{1,3}^{\ast }\right)
+\left( \varrho _{2,4}+\varrho _{2,4}^{\ast }\right) \right) \text{,} \qquad \\
\left\langle {\mathbf{S}}_{x}\right\rangle &=&\frac{1}{2}\left( \varrho
_{1,2}+\varrho _{2,1}+\varrho _{3,4}+\varrho _{4,3}\right) =\frac{1}{2}%
\left( \left( \varrho _{1,2}+\varrho _{1,2}^{\ast }\right) +\left( \varrho
_{3,4}+\varrho _{3,4}^{\ast }\right) \right) \text{,}
\end{eqnarray}
such as considering some properties of hermitian operators and performing some algebraic procedures to analyze the proper linear combination of Eq. (\ref{SpinSystemAXFirstOrder}), then the magnetization expressions are denoted by
\begin{eqnarray}
\left\langle \mathbf{I}_{x}\right\rangle  &=&\exp \left[ - \mathcal{C} \, \iota  _{f}^{\left( 1 \right) }\left(
t-t_{0}\right) \right] \left\langle \mathbf{I}_{x}\left( t_{0}\right)
\right\rangle _{f}+\exp \left[  - \mathcal{C} \, \iota  _{g}^{\left( 1 \right) }\left( t-t_{0}\right) \right]
\left\langle \mathbf{I}_{x}\left( t_{0}\right) \right\rangle _{g}\text{,} \quad \quad \\
\left\langle \mathbf{S}_{x}\right\rangle  &=&\exp \left[ - \mathcal{C} \, \iota  _{f}^{\left( 1 \right) }\left(
t-t_{0}\right) \right] \left\langle \mathbf{S}_{x}\left( t_{0}\right)
\right\rangle _{f}+\exp \left[  - \mathcal{C} \, \iota  _{g}^{\left( 1 \right) }\left( t-t_{0}\right) \right]
\left\langle \mathbf{S}_{x}\left( t_{0}\right) \right\rangle _{g}\text{,}
\end{eqnarray}
where the coefficients are resumed as follow
\begin{eqnarray*}
\left\langle \mathbf{I}_{x}\left( t_{0}\right) \right\rangle _{f} &=&-\frac{%
p_{3}p_{4}}{2p_{3}q_{4}-2p_{4}q_{3}}\left( \left( \varrho _{1,2}\left(
t_{0}\right) +\varrho _{1,2}^{\ast }\left( t_{0}\right) \right) +\left(
\varrho _{3,4}\left( t_{0}\right) +\varrho _{3,4}^{\ast }\left( t_{0}\right)
\right) \right)  \\
&&+\frac{p_{3}q_{4}}{2p_{3}q_{4}-2p_{4}q_{3}}\left( \left( \varrho
_{1,3}\left( t_{0}\right) +\varrho _{1,3}^{\ast }\left( t_{0}\right) \right)
+\left( \varrho _{2,4}\left( t_{0}\right) +\varrho _{2,4}^{\ast }\left(
t_{0}\right) \right) \right) \text{,}
\end{eqnarray*}%
\begin{eqnarray*}
\left\langle \mathbf{I}_{x}\left( t_{0}\right) \right\rangle _{g} &=&+\frac{%
p_{3}p_{4}}{2p_{3}q_{4}-2p_{4}q_{3}}\left( \left( \varrho _{1,2}\left(
t_{0}\right) +\varrho _{1,2}^{\ast }\left( t_{0}\right) \right) +\left(
\varrho _{3,4}\left( t_{0}\right) +\varrho _{3,4}^{\ast }\left( t_{0}\right)
\right) \right)  \\
&&-\frac{q_{3}p_{4}}{2p_{3}q_{4}-2p_{4}q_{3}}\left( \left( \varrho
_{1,3}\left( t_{0}\right) +\varrho _{1,3}^{\ast }\left( t_{0}\right) \right)
+\left( \varrho _{2,4}\left( t_{0}\right) +\varrho _{2,4}^{\ast }\left(
t_{0}\right) \right) \right) \text{,}
\end{eqnarray*}%
\begin{eqnarray*}
\left\langle \mathbf{S}_{x}\left( t_{0}\right) \right\rangle _{f} &=&-\frac{%
q_{3}p_{4}}{2p_{3}q_{4}-2p_{4}q_{3}}\left( \left( \varrho _{1,2}\left(
t_{0}\right) +\varrho _{1,2}^{\ast }\left( t_{0}\right) \right) +\left(
\varrho _{3,4}\left( t_{0}\right) +\varrho _{3,4}^{\ast }\left( t_{0}\right)
\right) \right)  \\
&&+\frac{q_{3}q_{4}}{2p_{3}q_{4}-2p_{4}q_{3}}\left( \left( \varrho
_{1,3}\left( t_{0}\right) +\varrho _{1,3}^{\ast }\left( t_{0}\right) \right)
+\left( \varrho _{2,4}\left( t_{0}\right) +\varrho _{2,4}^{\ast }\left(
t_{0}\right) \right) \right) \text{,} \\
\left\langle \mathbf{S}_{x}\left( t_{0}\right) \right\rangle _{g} &=&+\frac{%
p_{3}q_{4}}{2p_{3}q_{4}-2p_{4}q_{3}}\left( \left( \varrho _{1,2}\left(
t_{0}\right) +\varrho _{1,2}^{\ast }\left( t_{0}\right) \right) +\left(
\varrho _{3,4}\left( t_{0}\right) +\varrho _{3,4}^{\ast }\left( t_{0}\right)
\right) \right)  \\
&&-\frac{q_{3}q_{4}}{2p_{3}q_{4}-2p_{4}q_{3}}\left( \left( \varrho
_{1,3}\left( t_{0}\right) +\varrho _{1,3}^{\ast }\left( t_{0}\right) \right)
+\left( \varrho _{2,4}\left( t_{0}\right) +\varrho _{2,4}^{\ast }\left(
t_{0}\right) \right) \right) \text{,}
\end{eqnarray*}
the solutions found by  this  procedure behave under a bi-exponential evolution, and they work beyond  a single exponential dependence, as expected from previous theoretical studies  \cite{abragam1994Book,mcconnell1987Book}. The analysis of this result deserves a further discussion and it will not be addressed in this study.

\subsection{Density matrix dynamics: Time evolution of the elements}

\begin{SCfigure}
\includegraphics[width=3.0in]{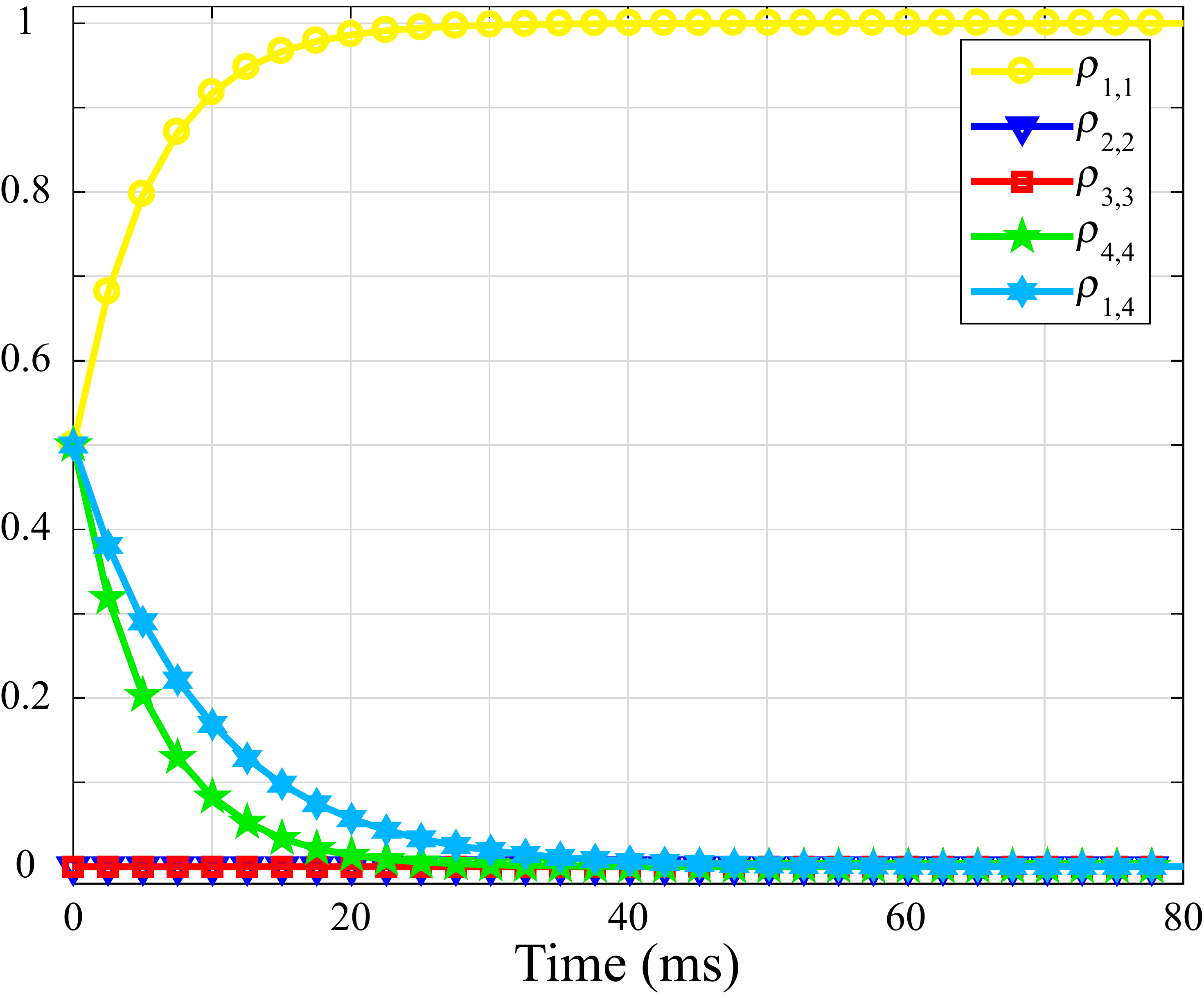} 
\caption{(Color online) The relaxation dynamics of five density matrix elements defined in Eq.(\ref{SpinSystemAXZeroOrder}) and Eq.(\ref{SpinSystemAXSecondOrderRho14}) are displayed, and each value is represented by symbols (with solid lines as a guide to the eyes). The evolution of $\varrho_{2,2}$, represented by the blue-triangle symbol, overlaps the dynamics of $\varrho_{3,3}$, represented by the red-square symbol. The evolution of $\varrho_{4,1}$ is not displayed.} \label{fig:DensidadeEvo}
\end{SCfigure}

Let us consider an initial quantum state  $ \vert  \psi (t_{0}) \rangle  \propto    \left\vert \uparrow \uparrow \right\rangle  + \left\vert \downarrow\downarrow \right\rangle   $ such that in the density matrix notation was represented  by
\begin{equation}
\varrho(t_{0})=\frac{(  \left\vert \uparrow \uparrow \right\rangle  + \left\vert \downarrow\downarrow \right\rangle)( \langle \uparrow\uparrow \vert + \langle \downarrow\downarrow \vert)}{2} \text{,} \label{PureStateInitial}
\end{equation}
and the steady-state represented by a pure quantum state  $ \vert  \psi^{\text{eq}} \rangle  =   \left\vert \uparrow \uparrow \right\rangle $, which can be  denoted in the density operator notation by
\begin{equation}
\varrho^{\text{eq}} = \vert \uparrow \uparrow  \rangle  \langle \uparrow \uparrow  \vert \text{.} \label{PureStateEquilibrio}
\end{equation}
We want to predict the time evolution of the initial quantum state and how it evolves to the steady-state using the solutions computed in Sec. \ref{sec:DescriptionSpinSystem}. Let us consider the  ``isotropic model'' for the spectral density functions defined in Eq. (\ref{ModelIsotropic}) and continue studying the anhydrous hydrofluoric acid but in a regime of the high strong static magnetic field at 9.4 Teslas. Therefore, the Larmor frequency of the Hydrogen and Fluorine nuclei are $\omega_{S}/2\pi =$ 400 MHz and $\omega_{I}/2\pi =$ 376 MHz, respectively; the correlation time $\tau=5\times 10^{-11}$ s, and the proportionality coefficient $\mathcal{C} = 3.7725 \times 10^{11} $Hz$^{2}$.  The density matrix elements prediction was devoted by applying Eq. (\ref{SpinSystemAXZeroOrderRho11}-\ref{SpinSystemAXZeroOrderRho44}) and Eq. (\ref{SpinSystemAXSecondOrderRho14}) because these assume non-null values. They are shown in Fig. \ref{fig:DensidadeEvo}. The non-null elements of the initial quantum state have the same intensity $\varrho_{1,1}\left(t_{0}\right)  =\varrho_{4,4}\left(t_{0}\right)=\varrho_{1,4}\left(t_{0}\right)=\varrho_{4,1}\left(t_{0}\right)=0.5$.
The element $\varrho_{1,1}$ evolves until the maximum value of the final steady-state, and the other ones evolve to the null value, as expected, showing that the solutions are working appropriately.

\begin{table}[t]
\begin{center}
\begin{tabular}{ccc|ccc} \hline\hline
\multicolumn{6}{c}{Violation of positivity} \\ \hline
 $^{1}$H - $^{19}$F & & $^{1}$H - $^{13}$C   & $^{1}$H - $^{19}$F  & & $^{1}$H - $^{13}$C  \\  \hline
 $ -0.53 \times 10^{-4}$ & &$ -0.43 \times 10^{-4}$  & $ -0.27 \times 10^{-2}$ & & $ -0.35 \times 10^{-1}$ \\ \hline\hline
 \multicolumn{3}{c|}{In-of-extreme narrow limit}    & \multicolumn{3}{c}{Out-of-extreme narrow limit} \\  \cline{1-3}\cline{4-6}
 \multicolumn{3}{c|}{$\tau= 5\times 10^{-11}$ s} & \multicolumn{3}{c}{$\tau= 5\times 10^{-10}$ s}  \\ \hline\hline
$\omega^{\text{C}}  \tau$ & $\omega^{\text{F}}  \tau$ & $\omega^{\text{H}}  \tau$  &$\omega^{\text{C}}  \tau$ &$\omega^{\text{F}}  \tau$ &  $\omega^{\text{H}}  \tau$    \\ \hline
 0.0316 & 0.1182 & 0.1257   & 0.3160 & 1.1820 & 1.2566    \\ \hline\hline
\end{tabular}%
\end{center}
\caption{At Left/Right were shown some parameter values to contextualize positivity violation assuming the In/Out extreme narrow limit regime, considering the isotropic model for the spectral density function. On top, the positive violation bounds for the pair  $^{1}$H - $^{19}$F and $^{1}$H - $^{13}$C. In the middle, the correlation time values used to establish the ``In'' and ``Out'' of extreme narrow limit. On the bottom, the factor used to characterize the extreme narrow limit.} \label{tab:ViolacaoPositividade}
\end{table}

Although the theoretical prediction displayed in Fig. \ref{fig:DensidadeEvo} visually preserves the positivity property by the Redfield equation, it is not general, and we would want to examine some bounds. Let us assume the high and low-temperature regimes to contextualize the discussion. The Eq. (\ref{QuantumStateEquilibrium}) represents the density matrix at the high-temperature regime, which describes a mixed state. This representation depends on the polarization parameter \cite{consuelo-leal2023JMR,dieguez2022,auccaise2011B,consuelo-leal2019,cius2022} or Boltzmann factor \cite{levitt2001Book,kowalewski2018Book}  frequently used in NMR, which is the ratio between the magnetic and thermal energies denoted by $\epsilon ^{S,I} = \frac{  \hbar \omega ^{S,I}}{k_{B}T}$. This parameter value would be between $ 10^{-6}$ and $10^{-4}$ for standard NMR spectrometers. Suppose this density matrix is transformed by performing any unitary propagator. In that case, any eigenvalue will be proportional to  $\frac{1}{\mathcal{Z}}\left( 1 \pm \frac{\epsilon}{2}\right)$, which will always be positive. On the other hand, at a very low-temperature regime, the density matrix represents pure states as in Eq. (\ref{PureStateInitial}) or (\ref{PureStateEquilibrio}). Monitoring the dynamics of each density matrix element, we identified that the element $\rho_{3,3}$ generates negative values. Considering the coupled system $^{1}$H - $^{19}$F and all the parameter values to generate Fig. \ref{fig:DensidadeEvo},  the lowest negative eigenvalue along the dynamic evolution is  $ -0.53 \times 10^{-4}$. If we change the nuclei species $^{19}$F to $^{13}$C and preserve all other parameters, the lowest negative eigenvalue is $ -0.43 \times 10^{-4}$. Those values violate the positivity property in both cases but are not critical. The picture changes if we increase the correlation time value to $\tau= 5\times 10^{-10}$s and perform the analysis for the coupled system  $^{1}$H - $^{19}$F, then the lowest negative eigenvalue is $ -0.27 \times 10^{-2}$. Similarly, for the coupled system  $^{1}$H - $^{13}$C, the lowest negative eigenvalue is $ -0.35 \times 10^{-1}$; these four values were detailed in Tab. \ref{tab:ViolacaoPositividade}. From these quantification procedures, the violation of the positivity has become more evident by increasing the correlation time, but also with the change of the nuclear species. Given this dependency, we figure out that the degree of positivity violation seems compatible with the extreme narrow limit. The quantification of the extreme narrow condition shows that for the case of $\tau= 5\times 10^{-11}$s, the factor  $\omega^{\text{H,C,F}} \ \tau < 1$ satisfies the condition (See values on Tab. \ref{tab:ViolacaoPositividade}), and for the case of   $\tau= 5\times 10^{-10}$s the factor  $\omega^{\text{C}} \ \tau < 1$ satisfies the condition and  $\omega^{\text{H,F}} \ \tau $ does not (See values on Tab. \ref{tab:ViolacaoPositividade}). Therefore, this brief analysis reveals an interesting qualitative relationship between the positivity property and the extreme narrow limit, which could be explored. Furthermore, the violation of positivity found at the low-temperature regime in this study is compatible with previous findings \cite{suarez1992,cheng2005,ishizaki2009}.

\section{Discussions}
\label{sec:Discussions}

In this study, the  Redfield superoperator characterizes the effects of the  bath acting on both spin particles that interact by the dipole-dipole coupling. Those effects are encoded by spectral density functions and correlation times of local field fluctuations  \cite{abragam1994Book,mcconnell1987Book}. An extended notation of Eq. (\ref{EquacaoRedfield}) is introduced in Eq. (\ref{EquationOfRelaxationDipolarHamiltonian04TermosComOperadorG}), showing an explicit representation of the  $\mathcal{R}_{\alpha\beta}^{ \alpha ^{\prime } \beta ^{\prime }}$ superoperator. The execution of those procedures allows to establish three superoperators for the three order coherences as denoted by $\boldsymbol{\mathcal{J}}^{\left( 0\right)}$,   $\boldsymbol{\mathcal{J}}^{\left( 1\right)}$, and  $\boldsymbol{\mathcal{J}}^{\left( 2\right)}$, explicitly detailed using matrix notation in Eq. (\ref{eq:SuperOperator0thOrder}),   (\ref{eq:SuperOperator1stOrder}), and   (\ref{eq:SuperOperator2ndOrder}), respectively. The eigenvalues of those superoperators $\iota _{\beta}^{\left( \alpha\right) }$ are resumed in Tab. \ref{tab:RelaxationRatesAX}, and they represent the relaxation rate constants denoted by $ R _{\beta}^{\left( \alpha\right) } =\mathcal{C} \ \iota _{\beta}^{\left( \alpha\right) }$, where $\mathcal{C}$ is a constant of proportionality as defined in Eq. (\ref{ConstantProportionalityGeneral}). Additionally, the density matrix elements solutions of the zero, first  and second-order coherences are resumed in Eq. (\ref{SpinSystemAXZeroOrder}), (\ref{SpinSystemAXFirstOrder}), and (\ref{SpinSystemAXSecondOrderRho14}), respectively. Those solutions obey a multiexponential decay, save the second-order coherence element. On the other hand, these definitions favor different applications.

Let us move back to Solomon's approach, particularly the experimental implementation using the anhydrous hydrofluoric acid and considering the dipole-dipole interaction between hydrogen and fluorine nuclei. In this study, we analyze  the dynamics of the longitudinal magnetization denoted by Eq. (44) of Ref. \cite{solomon1955}, which is the physical quantity monitored at three different initial conditions: \textit{(i)} a population saturation of the S specie, \textit{(ii)} a population inversion of the S specie, and \textit{(iii)} a simultaneous population inversion of both species. 

\begin{table}[b]
\begin{center}
\begin{tabular}{c|c|ll} \hline \hline
 \multicolumn{2}{c|}{Index} & \multicolumn{2}{c}{Superoperator eigenvalues} \\  \cline{1-2}
$\alpha$ & $\beta$ & \multicolumn{2}{c}{$  \iota _{\beta}^{\left( \alpha\right) }$}  \\ \hline
   & $a$ & $\frac{1}{6}\left( 3 J_{S}  +3 J_{I}  +4 J_{-}  \right) $ &  \\ 
   & $b$ & $\frac{3}{2}\left(  J_{S}  + J_{I}  \right) $ &  \\ 
 0 & $c$ & 0 &  \\ 
   & $e$ & 0 &  \\ 
   & $f$ & $\frac{J_{S}  + J_{I}}{2} +\frac{6 J_{+}  + J_{-}}{3} +\sqrt{\left( \frac{J_{S} - J_{I}}{2}  \right) ^{2}+\left( \frac{6 J_{+}  -  J_{-}}{3}  \right) ^{2}} $ &  \\ 
   & $g$ & $\frac{J_{S}  + J_{I}}{2} +\frac{6 J_{+}  + J_{-}}{3} -\sqrt{\left( \frac{J_{S} - J_{I}}{2}  \right) ^{2}+\left( \frac{6 J_{+}  -  J_{-}}{3}  \right) ^{2}} $ &  \\  \hline
   & $a$ & $\frac{2}{3}J_{0} +\frac{1}{4}  J_{S} +\frac{1}{4}  J_{I}  +\frac{1}{6} J_{-} + J_{+} +\sqrt{\left( \frac{J_{S}-J_{I}}{4}\right) ^{2}+\left( \frac{J_{-}+J_{0}}{3}\right) ^{2}}$ &  \\ 
 1 & $b$ & $\frac{2}{3}J_{0} +\frac{1}{4}  J_{S} +\frac{1}{4}  J_{I}  +\frac{1}{6} J_{-} + J_{+} -\sqrt{\left( \frac{J_{S}-J_{I}}{4}\right) ^{2}+\left( \frac{J_{-}+J_{0}}{3}\right) ^{2}}$ &  \\ 
   & $f$ & $\frac{2}{3}J_{0} +\frac{3}{4}  J_{S}  +\frac{3}{4} J_{I}  +\frac{1}{6}  J_{-} + J_{+} +\sqrt{\left( \frac{J_{S}-J_{I}}{4}\right) ^{2}+\left( \frac{J_{S}+J_{I}}{2}+\frac{J_{-}+J_{0}}{3}\right) ^{2}} $ &  \\ 
   & $g$ & $\frac{2}{3}J_{0} +\frac{3}{4}  J_{S}  +\frac{3}{4} J_{I} +\frac{1}{6} J_{-} + J_{+} -\sqrt{\left( \frac{J_{S}-J_{I}}{4}\right) ^{2}+\left( \frac{J_{S}+J_{I}}{2}+\frac{J_{-}+J_{0}}{3}\right) ^{2}}$ & \\ \hline
 2 & $a$ & $\frac{1}{2}\left( J_{S}  +J_{I} +4 J_{+}   \right) $ &  \\ \hline \hline
\end{tabular}
\end{center} \caption{Superoperator eigenvalues as a function of spectral density parameter for a zero, first, and second-order. } \label{tab:RelaxationRatesAX}
\end{table}

The population saturation technique performs the first experiment. In this description, the diagonal elements of the density matrix of the S specie must be equalized after a time $t_{0}$. The theoretical counterpart uses the density matrix notation, which is modeled by different approaches \cite{bull1992,viola2000,cius2022}, and   the diagonal density matrix elements are resumed in Eq. (\ref{QuantumStateSaturation}). After the saturation procedure finishes, the longitudinal magnetization dynamics of both nuclear species were modeled by the Redfield equation (a detailed explanation to demonstrate the solutions see Sec. \ref{sec:SaturationMagnetizationSNuclearSpecie}). Both dimensionless magnetizations for this saturation experiment are denoted by
\begin{eqnarray}
\Upsilon_{\text{Sat}}^{I} \left( t\right) &=&+\exp \left[ -\mathcal{C}J_{p}  t \right] \left( -2h_{1}\right) \sinh \left(\mathcal{C} J_{n}  t \right)  \text{,}  \label{MagnetizationZAdimenEspecieISaturationSnuclei} \\
\Upsilon_{\text{Sat}}^{S} \left( t\right) &=&-   \exp \left[ -\mathcal{C}J_{p}  t  \right]   \cosh \left( \mathcal{C} J_{n}  t  \right) +\exp \left[ -\mathcal{C}J_{p}  t  \right] \left(+ 2h_{2}\right)  \sinh \left(\mathcal{C} J_{n}  t  \right)    \text{,} \qquad  \label{MagnetizationZAdimenEspecieSSaturationSnuclei}
\end{eqnarray}
where the exponential and  hyperbolic functions depend on the spectral density functions defined by
\begin{eqnarray}
J_{p} &=&\frac{J_{S}  +J_{I}  }{2}+ \frac{6J_{+}  +J_{-} }{3}\text{,}  \label{SpectralDensityPositiveSignals} \\
J_{n} &=&\sqrt{\left( \frac{J_{S} -J_{I} }{2}\right) ^{2}+\left( \frac{6J_{+}  -J_{-} }{3}\right) ^{2}}\text{,} \label{SpectralDensityNegativeSignals}
\end{eqnarray}
and the magnetization depends on dimensionless coefficients $h_{1}$ introduced in Eq. (\ref{eq:CoefficientH1}) and $h_{2}$ in Eq. (\ref{eq:CoefficientH2}). These coefficients are a concise notation for a sophisticated dependence on spectral density functions detailed in  Eq. (\ref{eq:NormaTransParam}).

The population inversion technique performs the second experiment. In this setup, the population of the $S$ nuclei must be inverted after a time $t_{0}$. By theoretical counterpart, the initial density matrix is prepared through a standard $\pi$-pulse \cite{abragam1994Book},  and the diagonal density matrix elements are resumed in Eq. (\ref{QuantumStateInversion}). After the quantum state is initialized, the longitudinal magnetization evolution of both nuclear species is modeled by the Redfield equation (see Sec. \ref{sec:InversionMagnetizationSNuclearSpecie}). The magnetization results in twice the magnetization of the first experiment, and they are denoted by 
\begin{eqnarray}
\Upsilon_{\text{Inv}}^{I} \left( t\right) &=&+2 \Upsilon_{\text{Sat}}^{I} \left( t\right) \text{,}
\label{MagnetizationZAdimenEspecieIInversionSnuclei} \\
\Upsilon_{\text{Inv}}^{S} \left( t\right) &=&+2\Upsilon_{\text{Sat}}^{S} \left( t\right) \text{.}  \label{MagnetizationZAdimenEspecieSInversionSnuclei}
\end{eqnarray}

The third experiment performs the population inversion on both nuclei after $t_{0}$. The diagonal density matrix elements at the initial state are resumed in Eq. (\ref{QuantumStateInversionOnBoth}), and then both longitudinal magnetizations are modeled by the Redfield equation (see Sec.    \ref{sec:InversionMagnetizationBothNuclearSpecies}) and  denoted by
\begin{eqnarray}
\Upsilon_{\text{Inv-Both}}^{I} \left( t\right) &=&   -4\frac{\omega ^{I}h_{2}+\omega ^{S}h_{1}}{\omega ^{S} }\exp \left[ -\mathcal{C}J_{p}  t  \right]\sinh \left(\mathcal{C} J_{n}  t  \right)  \nonumber \\
& &-2\frac{\omega ^{I}}{\omega ^{S}}\exp \left[ -\mathcal{C}J_{p}  t  \right]\cosh \left(\mathcal{C} J_{n}  t  \right) \text{,} 
\label{MagnetizationZAdimenEspecieIBothInversion} \\
\Upsilon_{\text{Inv-Both}}^{S} \left( t\right) &=&    +4\frac{ \omega ^{S}h_{2}+\omega ^{I}h_{3}}{\omega ^{S}}\exp \left[ -\mathcal{C}J_{p}  t  \right]\sinh \left(\mathcal{C} J_{n}  t  \right) \nonumber \\
& &-2\exp \left[ -\mathcal{C}J_{p}  t  \right]\cosh \left(\mathcal{C} J_{n}  t  \right) \text{,}
\label{MagnetizationZAdimenEspecieSBothInversion}
\end{eqnarray}
where the parameter $h_{3}$ was introduced in Eq.  (\ref{eq:CoefficientH3}).

\begin{figure}[t]
\includegraphics[width=4.70in]{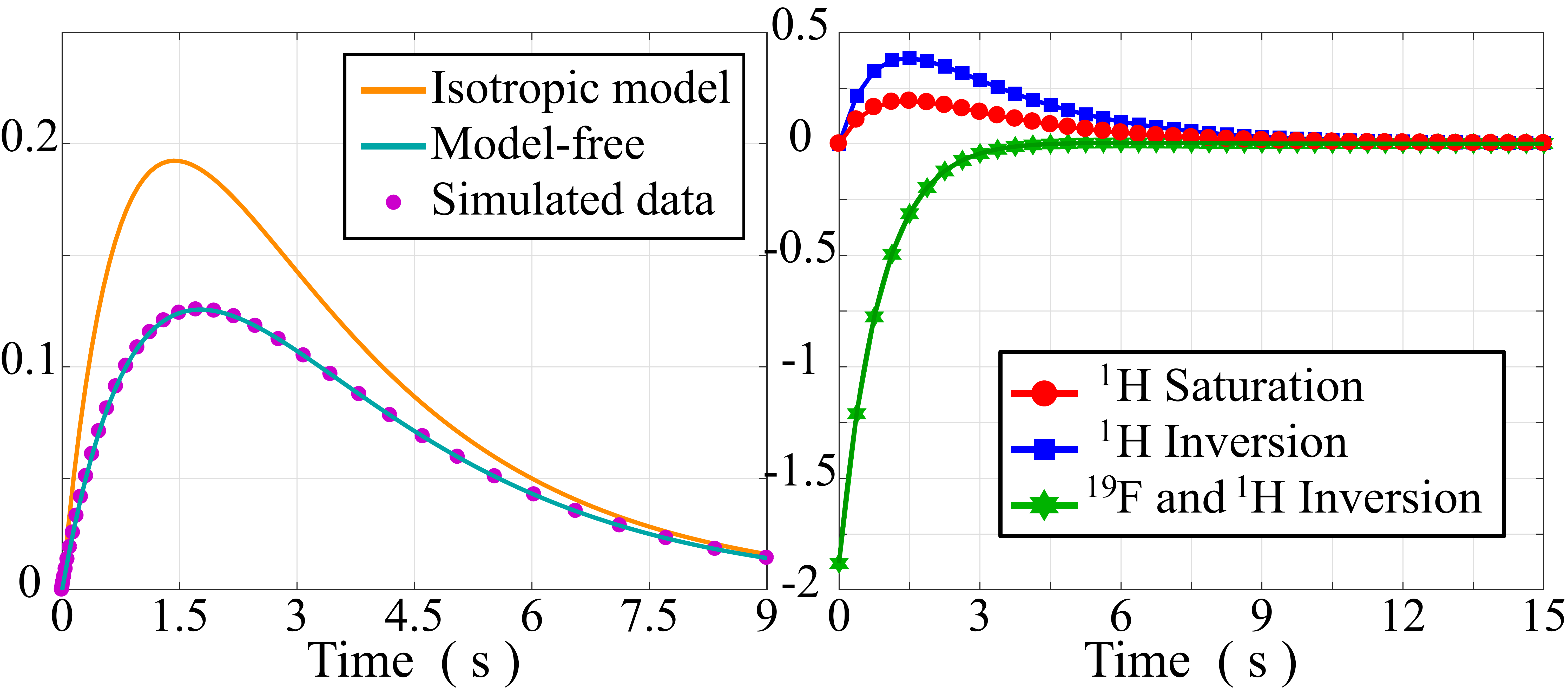}
\caption[Longitudinal magnetization]{(Color online) Longitudinal magnetization of the $I$ nuclear specie. On the left, the initial condition of the $S$ nuclear specie is established by the inversion population performed by a $\pi$-pulse \cite{abragam1994Book}. The quantum evolution representing the experimental results measured by I. Solomon (magenta dot symbols), theoretical prediction using the ``isotropic model'' (orange solid line) and ``model-free'' (dark green solid line) of spectral density functions using Eq. (\ref{FittingMagnetization}). On the right,  theoretical prediction using the ``isotropic model''  saturating the $S$ nuclear population of Eq. (\ref{MagnetizationZAdimenEspecieISaturationSnuclei}) (red circle symbols), inverting the $S$ nuclear population of Eq. (\ref{MagnetizationZAdimenEspecieIInversionSnuclei}) (blue square symbols) and inverting the population of both nuclear spin species of Eq. (\ref{MagnetizationZAdimenEspecieIBothInversion}) (green star symbols).} \label{fig:LongitudinalMagnetization}
\end{figure}

To verify the effectiveness of those magnetization solutions, we explore the experimental  setup introduced by Solomon \cite{bloembergen1948,solomon1955}, the anhydrous hydrofluoric acid at a  static magnetic field strength of $0.705$ T, as detailed in Ref. \cite{bloembergen1948}. Solomon highlighted two characteristic times $T_{1} = 1.27$s and $D_{1}=2.55$s. In this discussion, we use those time values to generate some data points (see magenta dots on the left side of Fig. \ref{fig:LongitudinalMagnetization}, and emulate the measurement of the Fluorine longitudinal magnetization.

Without loss of generality, we consider the Fluorine (Hydrogen) nuclei as the $I$ ($S$) species,  Eq. (\ref{MagnetizationZAdimenEspecieIInversionSnuclei}) must be rewritten to match the mathematical expression found in the caption of Fig. 4 of Ref. \cite{solomon1955}, denoted by
\begin{equation}
\Upsilon_{\text{Inv}}^{I*} \left( t\right)  =  -h_{1}\left(\exp \left[ -\mathcal{C} \ \iota _{g}^{\left( 0\right) } \   t \right]  - \exp \left[ -\mathcal{C} \ \iota _{f}^{\left( 0\right) }   \ t \right] \right)  \text{,} \label{FittingMagnetization}
\end{equation}
assuming an ``isotropic model'' for reorientation \cite{bull1972} encoded by the spectral density function  $J\left( \omega\right) =  2\tau_{c} /  ( 1+\left( \omega\tau_{c} \right)^{2} )$, where $\tau_{c}$ is the correlation time \cite{abragam1994Book}, and the proportionality parameter value is $\mathcal{C} = 47.9898 \times 10^{10} \ \text{Hz}^{2}$ using the distance value $r=96.098 \times 10^{-12} \ \text{m}$ between hydrogen-fluorine nuclei \cite{joutsuka2011}.   Therefore,  the correlation time computed   {to perform} the fitting procedure is $\tau_{c}= 0.2391 \times 10^{-12}$s, which is the best value. This value was used to predict the longitudinal magnetization, and it is displayed using a solid orange line on the left of Fig. \ref{fig:LongitudinalMagnetization}. From the figure, the solid line and data point visually show a mismatch between the theoretical prediction and the simulated data. This was expected because the anhydrous hydrofluoric acid does not relax with a pure dipole-dipole coupling, but mixing dipole-dipole and exchange interactions \cite{solomon1956}. However, considering the ``Model-free'' introduced by Lipari-Szabo in  Eq. (1) of Ref. \cite{lipari1982A}, the simulated data can be fitted with  great accuracy, as shown by the dark green solid line on the left of Fig. \ref{fig:LongitudinalMagnetization}, with correlation time values $\tau_{M}=31.99704 \times 10^{-9} $s, $\tau_{e}=0.80751 \times 10^{-12} $s and order parameter $S=0.005932$.  The second fitting procedure does not imply any physical interpretation of the molecule dynamics generated by the model. It only points out that the assumption of the sources of fluctuations can be considered by two ones. This remark is compatible with an extended description introduced by I. Solomon and N. Bloembergen, which changed this misinterpretation \cite{solomon1956}.

Continuing to explore the relaxation rate constants as resumed in Tab. \ref{tab:RelaxationRatesAX}, if we consider the ``isotropic model'' for the spectral density functions and at the extreme narrow limit, then  the spectral density functions satisfy   $J_{I}  \ \approx \  J_{S}  \ \approx \  J_{+}  \ \approx \  J_{-}  \ \approx 2 \tau_{c}= 0.4782  \times 10^{-12} $s. Analyzing both  Eq. (\ref{SpectralDensityPositiveSignals}) and (\ref{SpectralDensityNegativeSignals}) and considering  $J_{S} \approx  J_{I}$, it can be shown that
\begin{eqnarray}
 J_{p} &\approx & \left( 3J_{I} +  6J_{+}  +J_{-}\right) / 3  =  20 \tau_{c}/3   = 1.5939 \, \text{ps,} \ \ \\
 J_{n} &\approx & \left( 6J_{+}  -J_{-}\right) /3  =  10 \tau_{c}/3 =0.7970 \, \text{ps,}
 \end{eqnarray}
where the ratio between them satisfies  $J_{n}/J_{p} = 1/2 $, both expressions may better describe the coefficients  $\rho$ and $\sigma$, as denoted in Eq. (15) of Ref. \cite{solomon1955} or  Eq. (22) of Ref. \cite{hausser1968}.  Also, the definition of both parameters favors  us to identify the coupling parameter $\xi$  introduced by A. Abragam \cite{hausser1968,abragam1955} or  the nuclear Overhauser enhancement factor $\eta_{I}(S)$ as was defined in Eq. (10) of Ref. \cite{freeman1974}.

Other key characteristics of the relaxation model highlighted by I. Solomon stays about the  transition probability ratios denoted by $w_{0}:w_{1}:w_{2}=2:3:12$ (see comment on page $6$ of Ref. \cite{muller-warmuth1983}). The same  characteristic can be played if Eq. (\ref{SpectralDensityPositiveSignals}) is  rewritten as  $6J_{p} = 2J_{-}+3J_{I}+3J_{S}+12J_{+} $, where each constant coefficient is proportional to each transition probability. A similar property is stressed analyzing Eq. (\ref{SpectralDensityNegativeSignals}).

The computation of those parameter values must be applied to predict the  longitudinal magnetization dynamics of the $I$ nuclear specie   at the three experimental setups, which are displayed on the right of Fig. \ref{fig:LongitudinalMagnetization}. The saturation (red circle symbols) and the population inversion (blue square symbols) of the $S$ nuclear species experiments highlight the enhancement effect on the magnetization of the $I$ specie. In contrast, the population inversion of both species (green star symbols) reveals the standard dynamics of the inversion-recovery experiment \cite{abragam1994Book}. Those predictions evidence the efficiency of the enhancement effect by twice times the longitudinal magnetization  of the population inversion (blue square symbols) against the population saturation (red circle symbols) experiment. The longitudinal magnetization of the $S$ nuclear specie at the three experimental setups was computed, and those results reveal the standard dynamics of the inversion-recovery experiment \cite{abragam1994Book}, but they are not shown.  

Additionally, the analytical approach introduced in this study helps to determine the  precise time $t_{\text{m}}$ of enhancement to achieve the maximum magnetization value. This maximum can be found by differentiating Eq.  (\ref{MagnetizationZAdimenEspecieISaturationSnuclei})  or (\ref{MagnetizationZAdimenEspecieIInversionSnuclei}) and equating it to zero, $ \partial\left(\Upsilon_{\text{Inv}}^{I} \left( t\right)\right)/\partial t =0$. Therefore, the maximum magnetization happens at
\begin{equation}
t_{\text{m}}=\frac{1}{\mathcal{C}J_{n}}\tanh ^{-1}\left( \frac{J_{n}}{J_{p}}\right) \label{eq:TempoMaximoGanho} \text{,}
\end{equation}
and for the data of Fig. \ref{fig:LongitudinalMagnetization}  is $t_{\text{m}}= 1.4363 $s.

If the ``Model-free'' is considered and  an analysis of Eq. (\ref{SpectralDensityPositiveSignals}) and (\ref{SpectralDensityNegativeSignals}) at the extremely  narrow limit is devoted, then both expressions are quantified
\begin{eqnarray}
 J_{p} &\approx & \left( 3J_{I} +  6J_{+}  +J_{-}\right) / 3    \approx 1.22897 \, \text{ps,} \ \ \\
 J_{n} &\approx & \left( 6J_{+}  -J_{-}\right) /3  \approx  0.41180 \, \text{ps,}
\end{eqnarray}
where the ratio between them satisfies  $J_{n}/J_{p} \approx 0.3351 $, it implies on characteristic times $T1 = 1.27$ s and $D1 = 2.55$ s as measured by I. Solomon \cite{solomon1955}. Moreover, the maximum magnetization computed using Eq. (\ref{eq:TempoMaximoGanho}) happens at  $t_{\text{m}}= 1.7637 $s.

Furthermore, I. Solomon claimed a mathematical disagreement between the theoretical approach, as specified by Eq. (39) and (40), and the experimental results of Ref. \cite{solomon1955}. The solutions introduced by the Redfield theory match the value he predicted regarding  the coefficients $h_{1}$ and $h_{2}$.  Moreover, at the extremely narrow limit, it assumes  the value $-h_{1}=1/2$ and $h_{2}=0$, even for the ``isotropic model'' or ``model-free'' of the spectral density function. On other regimes, both coefficients assume different values at different correlation times and for other spectral density functions because they cover a wide range of vibrational movements representing the action of an environment. To put in evidence, the versatility of the solutions generated by computing the Redfield equation, in Fig. \ref{fig:AuxiliaryFunctionsForMagnetization}, were shown different range values of $h_{k}$ coefficients at different correlation times and assuming the ``isotropic model''. At each different model assumed then, different range values of $h_{k}$ coefficients will be generated (data not shown but easily verified using Eq. (\ref{eq:NormaTransParam})).

Finally, the use of these solutions can be implemented in quantum information processing, performing predictions on quantum protocols \cite{dieguez2022,consuelo-leal2019}, modeling the non-classical correlations of quantum states resilient to environment effects \cite{auccaise2011B}, and the evolution of each density matrix element of long-live quantum states \cite{emondts2014}, helping to distinguish between experimental errors and environment effects.
 
\section{Conclusions}
\label{sec:Conclusions}

In the present analysis, we introduced a theoretical study about the relaxation dynamics of an unlike spin pair system interacting by a pure dipole-dipole coupling. The analytical solution of the Redfield equation was reported, and mathematical expressions for each density matrix element were found. Those solutions were applied to describe the longitudinal magnetization dynamics. The coefficients found by performing the density matrix analysis of the Redfield equation saved the discrepancies reported by I. Solomon.

It is well known that Solomon's analysis drew out some other experimental setups, and with the analytical solutions introduced in this study, new insights will be possible. Indeed, we mainly focused on the discussion of Solomon's approach because it was the starting point to explore another physical context, such as Dynamic Nuclear Polarization or even to Nitrogen-vacancy center formalism to cite some of them.

\section{Acknowledgements}
H.D.F.S. is supported by the CNPq-Brazil grant 406621/2021-7 and FAPEMIG-Brazil grant APQ-00782-21.
R.A. acknowledges the financial support of CNPq (309023/2014-9, 459134/2014-0) and CNPq INCT-IQ (465409/2014-0). The authors would want to thank the anonymous referee for his/her valuable comments, which helped us to improve the paper.

\section{Conflict of interest }
All authors declare that they have no conflicts of interest.

\section{Data availability} The datasets generated during and/or analyzed during the current study are available from the corresponding author on reasonable request.

\begin{appendices}

\section{Second-rank operators}

The matrix representation of the second-rank operators denoted in Eq.  (\ref{eq:DipolarTensorOperatorRank2}) 
\begin{equation*}
{\mathbf{A}}^{\left( -2\right) }=\left[ 
\begin{array}{cccc}
0 & 0 & 0 & 0 \\ 
0 & 0 & 0 & 0 \\ 
0 & 0 & 0 & 0 \\ 
1 & 0 & 0 & 0%
\end{array}%
\right] \text{,\qquad } {\mathbf{A}}^{\left( -1\right) }=\frac{1}{2}%
\left[ 
\begin{array}{cccc}
0 & 0 & 0 & 0 \\ 
1 & 0 & 0 & 0 \\ 
1 & 0 & 0 & 0 \\ 
0 & -1 & -1 & 0%
\end{array}%
\right] \text{,\qquad } {\mathbf{A}}^{\left( 0\right) }=\frac{1}{%
\sqrt{6}}\left[ 
\begin{array}{cccc}
1 & 0 & 0 & 0 \\ 
0 & -1 & -1 & 0 \\ 
0 & -1 & -1 & 0 \\ 
0 & 0 & 0 & 1%
\end{array}%
\right] \text{,}
\end{equation*}
\begin{equation}
{\mathbf{A}}^{\left( +1\right) }=\frac{1}{2}%
\left[ 
\begin{array}{cccc}
0 & -1 & -1 & 0 \\ 
0 & 0 & 0 & 1 \\ 
0 & 0 & 0 & 1 \\ 
0 & 0 & 0 & 0%
\end{array}%
\right] \text{,\qquad } {\mathbf{A}}^{\left( +2\right) }=\left[ 
\begin{array}{cccc}
0 & 0 & 0 & 1 \\ 
0 & 0 & 0 & 0 \\ 
0 & 0 & 0 & 0 \\ 
0 & 0 & 0 & 0%
\end{array}%
\right] \text{.}
\end{equation}
This set of five matrices was used to apply in the Eq. (\ref{EquationOfRelaxationDipolarHamiltonian04TermosComOperadorG}) to construct the linear systems of Eq. (\ref{SpinSystemAXZeroOrder}), (\ref{SpinSystemAXFirstOrder}), and (\ref{SpinSystemAXSecondOrderRho14}).

\end{appendices}





    \end{document}